\documentclass[aps,prd,preprintnumbers,groupedaddress,nofootinbib,eqsecnum,notitlepage,dvipdfmx,10pt]{revtex4-1}
\usepackage{here}
\usepackage{graphicx}
\usepackage{amsmath,amsthm,amssymb}
\usepackage{bm}
\usepackage{color}

\usepackage{amsfonts}
\usepackage{dcolumn}
\usepackage{color}
\usepackage{hyperref}
\allowdisplaybreaks[1]

\begin{document}
\newcommand{\newc}{\newcommand}
\newcommand{\rd}{{\rm d}}
\newcommand{\bA}{\bar{A}}

\newcommand{\rk}[1]{{\color{red} #1}}
\newcommand{\ben}{\begin{eqnarray}}
\newcommand{\een}{\end{eqnarray}}
\newc{\be}{\begin{equation}}
\newc{\ee}{\end{equation}}
\newc{\ba}{\begin{eqnarray}}
\newc{\ea}{\end{eqnarray}}
\newc{\bea}{\begin{eqnarray*}}
\newc{\eea}{\end{eqnarray*}}
\newc{\D}{\partial}
\newc{\ie}{{\it i.e.} }
\newc{\eg}{{\it e.g.} }
\newc{\etc}{{\it etc.} }
\newc{\etal}{{\it et al.}}
\newcommand{\nn}{\nonumber}
\newc{\ra}{\rightarrow}
\newc{\lra}{\leftrightarrow}
\newc{\lsim}{\buildrel{<}\over{\sim}}
\newc{\gsim}{\buildrel{>}\over{\sim}}
\newc{\aP}{\alpha_{\rm P}}

\makeatother

\title{
Neutron stars with a generalized Proca hair and spontaneous vectorization}

\vspace{0.5cm}

\author{
Ryotaro Kase$^{1}$,
Masato Minamitsuji$^{2}$, and 
Shinji Tsujikawa$^{1}$}

\affiliation{
$^1$Department of Physics, Faculty of Science, Tokyo University of Science, 1-3, Kagurazaka,
Shinjuku-ku, Tokyo 162-8601, Japan\\
$^2$Centro Multidisciplinar de Astrofisica - CENTRA, Departamento 
de Fisica, Instituto Superior Tecnico - IST, Universidade de Lisboa - 
UL, Avenida Rovisco Pais 1, 1049-001 Lisboa, Portugal}

\begin{abstract}
In a class of generalized Proca theories, we study the existence of neutron star solutions with a nonvanishing temporal component of the vector field $A_\mu$ approaching 0 toward spatial infinity,
as they may be the endpoints of tachyonic instabilities of neutron star solutions 
in general relativity with $A_{\mu}=0$.
Such a phenomenon is called spontaneous vectorization, which is analogous to spontaneous scalarization in scalar-tensor theories with nonminimal couplings to the curvature or matter. 
For the nonminimal coupling $\beta X R$, where $\beta$ is a coupling 
constant and $X=-A_{\mu}A^{\mu}/2$, we show that there exist both 0-node and 1-node vector-field solutions, irrespective of the choice of the equations of state of nuclear matter. The 0-node solution, which is present only for $\beta=-{\cal O}(0.1)$, may be induced 
by some nonlinear effects such as the selected choice of initial conditions. 
The 1-node solution exists for $\beta=-{\cal O}(1)$, which suddenly emerges above a critical  central density of star and approaches the general relativistic branch with the increasing central density. 
We compute the mass $M$ and radius $r_s$ of neutron stars for some realistic equations of state and show that the $M$-$r_s$ relations of 0-node and 1-node solutions exhibit notable difference from those of scalarized solutions in scalar-tensor theories. 
Finally, we discuss the possible endpoints of tachyonic instabilities.
\end{abstract}

\date{\today}


\maketitle

\section{Introduction}
\label{introsec}

The advent of gravitational-wave (GW) astronomies \cite{Abbott2016}, 
along with observations of binary pulsars \cite{binary}, 
opened up a new channel for probing physics of high-density matter inside 
neutron stars (NSs). 
In terms of the tidal deformability of NSs, the data of 
the GW170817 event \cite{GW170817} put constraints on the relation 
between mass and radius of NSs.
In addition, the recent X-ray data collected by NASA's {\it NICER} mission \cite{NICER1} 
inferred mass and radius of a millisecond pulsar as well as equation of state (EOS) \cite{NICER2}. 
With the accumulation of GW and other events 
in the future along with the increasing
accuracy of measurements, we will be able to place tighter bounds 
on NS EOSs as well as the possible deviation 
from General Relativity (GR) in high-density regions. 
In particular, whether or not some extra degrees of freedom beyond 
GR and standard model of particle physics exist around strong 
gravitational objects is an important question. 
Moreover, such new degrees of freedom may be related to the problems 
of dark sectors in our Universe such as dark matter and dark energy 
(see, e.g., Ref.~\cite{Baker:2017hug} for the observational constraint 
on dark energy models from GW170817 event).

One of the simplest and well-motivated modifications to GR 
in the regime of strong gravity is to introduce nonlinear 
scalar curvature terms like $R^2$ in the Lagrangian, besides the 
Einstein-Hilbert term $R$ \cite{Staro}. 
This theory, which is known as a class of $f(R)$ gravity \cite{Bergmann,Ruz}, 
has one additional scalar degree of freedom in comparison to GR, 
with an effective potential arising from the gravitational origin \cite{fRreview}. 
NS solutions in $f(R)$ gravity have been extensively studied in the 
literature \cite{Cooney:2009rr,Arapoglu:2010rz,Orellana:2013gn,Astashenok:2013vza,Yazadjiev:2014cza,Resco:2016upv}, 
but the presence of nonvanishing scalar mass can give rise to an exponential 
growing mode outside the star \cite{Kase:2019dqc}. This is the case for the model 
$f(R)=R+\alpha R^2$ with $\alpha$ being constant, in which
the scalar degree of freedom does not 
vanish at spatial infinity unless EOS inside NSs is chosen in a specific way \cite{Ganguly:2013taa}. 

One can express $f(R)$ theories in terms of the action of scalar-tensor theories
with a nonminimal coupling to the Ricci scalar.
There are also other scalar-tensor theories in which 
a scalar field $\phi$ is nonminimally coupled to the Ricci 
scalar $R$ of the form $F(\phi)R$, where $F(\phi)$ is a function 
of $\phi$ \cite{Brans,Fujii}.
For the massless scalar field in Brans-Dicke theories 
with the coupling $F(\phi)=e^{-2Q\phi/M_{\rm pl}}$ \cite{Brans}, 
where $Q$ is a constant and $M_{\rm pl}$ is the reduced Planck mass, 
it is possible to realize a nontrivial configuration of the field inside 
NS with $\phi$ approaching 0 at spatial infinity \cite{Kase:2019dqc}. 
In scalar-tensor theories, the construction of hairy solutions in compact 
objects and their observational signatures have been studied 
in Refs.~\cite{Kobayashi:2018xvr,Saltas:2019ius,Babichev:2016jom}.

While the theory with a monotonic coupling function $F(\phi)$
only admits NS solutions with a nontrivial profile of $\phi$,
a theory with $F_{,\phi}(0)=0$, where $F_{,\phi}={\rm d}F/{\rm d}\phi$,
also admits NS solutions in GR with $\phi=0$.
The effective mass squared for small perturbations about the GR NS solution
is given by 
$m_{\rm eff}^2=- (M_{\rm pl}^2/2)[F_{,\phi \phi}(0)/\omega (0)]R$, 
see Eq.~\eqref{Jaction} in Appendix for our convention.
Provided that $F_{,\phi \phi}(0)>0$ with $R>0$ and $\omega (0)>0$,
there is a tachyonic instability of the GR branch, which can be triggered by  
spontaneous growth of $\phi$ toward the other nontrivial branch.
Then, NSs may eventually acquire a scalar hair,
whose phenomenon is dubbed {\it spontaneous scalarization}.
Spontaneous scalarization is particularly interesting,
as it would modify the gravitational interaction only 
in strong-gravity regimes and be directly tested via future GW measurements.

Damour and Esposito-Farese \cite{Damour,Damour2} proposed a concrete nonminimal coupling $F(\phi)=e^{-\beta \phi^2/(2M_{\rm pl}^2)}$ for spontaneous scalarization, 
which satisfies the conditions $F_{,\phi}(0)=0$ and
$F_{,\phi \phi}(0)>0$ for $\beta<0$.
They showed that there exist 
the spherically symmetric and static NS solutions with a nonvanishing field 
configuration besides the GR branch.
In Refs.~\cite{Harada:1998ge,Novak:1998rk,Silva:2014fca},
it was shown that 
the GR solution can be unstable to trigger 
spontaneous scalarization to the other nontrivial branch
for $\beta<-4.35$, 
depending very weakly on the choice of EOSs. 
Since the scalarized solution has a scalar charge associated with 
the energy loss through dipolar radiation, 
binary-pulsar
observations have placed the bound $\beta>-4.5$
\cite{Freire:2012mg}
(see also Refs.~\cite{binary1,binary2,binary3}).
These two results confine the coupling $\beta$ to a limited range.
On the other hand, it has been shown that 
spontaneous scalarization can also be realized 
for black holes,
in the presence of couplings to the Gauss-Bonnet 
term \cite{Kleihaus:2015aje,Doneva:2017bvd,Silva:2017uqg,Antoniou:2017acq,Antoniou:2017hxj,Minamitsuji:2018xde,Cunha}
and to the electromagnetic field \cite{Stefanov,Herdeiro1,Herdeiro2,Herdeiro3,Ikeda}.
The threshold values of generalized scalar-tensor couplings for the tachyonic instability and the onset of spontaneous scalarization have been extensively studied in Ref.~\cite{Ventagli:2020rnx}.

The nonminimally coupled scalar field is not only the possibility 
for modifying the physical property of NSs, but the vector field 
$A_{\mu}$ coupled to gravity should also affect the structure of NSs.
Thus, in analogy to spontaneous scalarization, it is of interest to 
study the possibility of {\it spontaneous vectorization} in 
vector-tensor theories.
The standard Einstein-Maxwell theory is given by the Lagrangian 
$L=M_{\rm pl}^2R/2-F_{\mu \nu}F^{\mu \nu}/4$, where 
$F_{\mu \nu}=\nabla_{\mu}A_{\nu}-\nabla_{\nu}A_{\mu}$ 
is the Maxwell tensor with $\nabla_{\mu}$ being the covariant 
derivative operator.
The simple example for modifying the gravitational interaction in GR is to 
introduce the nonminimal coupling $G_4(X)R$, where $G_4$ is a function 
of $X=-A^{\mu}A_{\mu}/2$. In this Hellings and Nordtvedt theory \cite{Hellings}, 
Annulli {\it et al.} \cite{Annulli} 
found NS solutions with a nonvanishing temporal vector component 
approaching $0$ at spatial infinity, besides the GR branch.
The possibility of spontaneous vectorization was analyzed 
from the viewpoint of Einstein frame
in Refs.~\cite{Rama1,Rama2}.

The vector-tensor theories with the coupling $G_4(X)R$ alone generally give
rise to derivatives higher than second order in the field equations of motion, 
so it can be prone to the problem of Ostrogradski 
instabilities \cite{Ostro,Ostro2} with the Hamiltonian unbounded 
from below. The theories remain up to second order by taking into 
account additional derivative-interaction terms to the Lagrangian of the form 
$G_{4,X}(X) [ (\nabla_{\mu} A^{\mu})^{2}-\nabla_{\mu} A_{\nu}
\nabla^{\nu} A^{\mu}]$, where $G_{4,X} \equiv 
{\rm d} G_4/{\rm d} X$. 
They are known as a class of generalized Proca (GP) 
theories \cite{Heisenberg,Tasinato,Allys,Jimenez2016}, in which 
the $U(1)$ gauge symmetry is broken by the explicit 
$X$ dependence in $G_4$. 
The application of GP theories to the late-time cosmic 
acceleration \cite{DeFelice:2016yws,DeFelice:2016uil,deFelice:2017paw,Nakamura:2018oyy} 
and the screening of fifth forces \cite{Vain1,Vain2} around local objects
on the weak gravitational background has been widely 
studied in the literature.

If we apply GP theories to compact objects on the strong gravitational background, 
there are hairy NS and black hole solutions 
with nontrivial vector-field 
profiles \cite{Tasinato1,Tasinato2,Minamitsuji,GPBH,GPBH2,Fan,Cisterna,Babichev17,KMT17}. 
In Refs.~\cite{Tasinato2,KMT17}, NS solutions were 
studied for polytropic EOS for the models which only have hairy NS solutions.
Models associated with spontaneous vectorization should possess
a nonvanishing vector field with $A_{\mu}$ approaching 0 
at spatial infinity, which hence would arise from a tachyonic instability of 
the GR  solution with $A_{\mu}=0$.
For the former branch, there will be a nontrivial modification 
to the structure of NSs (like mass and radius) 
with $A_{\mu}$ carrying a vector charge. 

In this paper, we study NS solutions in 
the above class of GP theories 
with the nonvanishing vector-field profile approaching 0 
far outside the star.
To describe realistic nuclear interactions inside NSs, 
we use the analytic representations of
SLy and BSk20 EOSs 
given in Refs.~\cite{Haensel:2004nu,Potekhin:2013qqa}.
We consider the simple nonminimal coupling $\beta X$ in $G_4(X)$, 
which allows for nontrivial vector-field solutions 
besides the GR branch. 
We will show that there are either 0-node or 1-node solutions 
depending on whether the coupling $\beta$ is of order 
$-0.1$ or $-1$, respectively. 
The 0-node solution in GP theories has a different property compared to 
that in scalar-tensor theories, in that the former may 
appear through some nonlinear effects like the selected choice of 
initial conditions. 
We will also show that the value of $\beta$ 
for the 1-node solution 
is consistent with the onset of instability.
We will compute the mass and radius of NSs for both 0-node and 
1-node solutions and investigate how they are modified from 
those in GR. Finally, we will clarify the difference from the case of 
scalarized solutions and discuss the possible endpoints of 
tachyonic instabilities.

Throughout the paper, we use the natural units $c=\hbar=1$, 
where $c$ is the speed of light and $\hbar$ is 
reduced Planck constant. When these fundamental constants are 
needed in numerical computations, we recover them and adopt 
their concrete values 
$c = 2.9979 \times 10^{10}$ cm $\cdot$ s$^{-1}$ 
and $\hbar=1.0546 \times 10^{-27}$~erg $\cdot$ s, and 
the Newton gravitational constant 
$G=6.6743 \times 10^{-8}$~g$^{-1} 
\cdot$~cm$^3\cdot$ s$^{-2}$. 
In terms of the normalization of vector field, it is convenient 
to use the reduced Planck mass $M_{\rm pl}$, 
which is related to $G$ as $M_{\rm pl}=(8 \pi G)^{-1/2}$.

\section{Generalized Proca theories and relativistic stars}
\label{setuosec}

We consider a vector field $A_{\mu}$ breaking the $U(1)$ 
gauge symmetry due to the presence of nonminimal coupling 
$G_4(X)R$, where $X$ is a function of $X=-A^{\mu} A_{\mu}/2$ 
and $R$ is the Ricci scalar. 
This type of nonminimal vector coupling to gravity was first introduced 
by Hellings and Nordtvedt in 1973 \cite{Hellings}, 
but we need to worry for Ostrogradski instabilities associated 
with the existence of derivatives higher than second order.
The theory can be made second order 
by adding a counter term that eliminates higher-order derivatives. 
The action of such second-order GP theories is given 
by \cite{Heisenberg,Tasinato,Allys,Jimenez2016}
\be
{\cal S}=\int {\rm d}^{4}x \sqrt{-g} 
\left[ G_{4}(X) R + G_{4,X}(X) \left\{ (\nabla_{\mu} A^{\mu})^{2} 
-  \nabla_{\mu} A_{\nu} \nabla^{\nu} A^{\mu}\right\}
-\frac{1}{4}F_{\mu \nu}F^{\mu \nu} \right]
+{\cal S}_m (g_{\mu \nu}, \Psi_m)\,,
\label{action}
\ee
where $g$ is the determinant of metric tensor.
We take into account the action ${\cal S}_m$ of 
matter fields $\Psi_m$, which are assumed to be minimally coupled 
to gravity. The action (\ref{action}) can be generalized further to 
include other derivative and nonminimal 
couplings \cite{Heisenberg,Tasinato,Allys,Jimenez2016,HKT16,KNY16}, 
but we will focus on the theory (\ref{action}) for simplicity.

\subsection{Background equations}

The line element on a spherically symmetric and static 
background is given by 
\be
{\rm d}s^{2} =-f(r)
{\rm d}t^{2} +h^{-1}(r) {\rm d}r^{2} + 
r^{2} \left( {\rm d}\theta^{2}+\sin^{2}\theta\, 
{\rm d} \varphi^{2} \right)\,,
\label{metric}
\ee
where $f$ and $h$ are functions of the radial coordinate 
$r$ from the center of symmetry.
On this background, the vector field is 
expressed in the form 
\be
A_{\mu}=\left( A_0(r), A_1(r), 0, 0 \right)\,,
\label{vector_ansatz}
\ee
where $A_0(r)$ and $A_1(r)$ correspond to temporal and 
radial components, respectively, which depend on $r$ alone. 
Then, the quantity $X$ is expressed as 
\be
X=\frac{A_0^2}{2f}-\frac{h A_1^2}{2}\,.
\ee
For the matter sector, we consider a single perfect fluid whose 
mixed energy-momentum tensor is given by 
$T^{\mu}_{\nu}={\rm diag} (-\rho(r),P(r),P(r),P(r))$, where 
$\rho(r)$ and $P(r)$ are the density and pressure, respectively.
{}From the matter continuity equation $\nabla_{\mu} T^{\mu}_{\nu}=0$, 
we obtain
\be
P'+\frac{f'}{2f} \left( \rho+P \right)=0\,,
\label{Peq}
\ee
where a prime represents a derivative with respect to $r$.

Variation of the action (\ref{action}) with respect to 
$A_1$ leads to 
\be
A_1 \left[ \left(f-fh-rh f' \right)f G_{4,X}
+\left\{ fh (rf'+f) A_1^2-rA_0 (A_0 f' -2f A_0') 
\right\} h G_{4,XX} \right]=0\,.
\ee
This shows that there exists the branch satisfying  
\be
A_1=0\,.
\ee
Throughout this paper, we will focus on this branch. 
The difference from the solution in GR arises from 
the temporal component $A_0$.
Varying the action (\ref{action}) with respect to $f$, $h$, 
and $A_0$, it follows that 
\ba
& & h' = \frac{4(1-h)(G_4 f-A_0^2 G_{4,X})
-2\rho r^2 f-r^2 h A_0'^2}
{4r (G_4f-A_0^2 G_{4,X})}\,,
\label{be1} \\
& & f' = \frac{f[4f(1-h)G_4+2Pr^2f 
-rh A_0' (rA_0'+8A_0 G_{4,X})]}
{4rh(G_4 f-A_0^2 G_{4,X})}\,,
\label{be2} \\
& &
A_0''+\left( \frac{2}{r}-\frac{f'}{2f}+\frac{h'}{2h} \right)
A_0'+\frac{2}{r^2 h} G_{4,X} \left( rh'+h-1 \right) A_0=0\,.
\label{be3}
\ea
Substituting Eq.~(\ref{be1}) into Eq.~(\ref{be3}), the temporal 
vector component obeys 
\be
A_0''+\left( \frac{2}{r}-\frac{f'}{2f}+\frac{h'}{2h} \right)
A_0'-\frac{G_{4,X}(2f \rho+h A_0'^2)}
{2h(G_4 f-A_0^2 G_{4,X})}A_0=0\,.
\label{Aeq}
\ee
Provided that EOS $P=P(\rho)$ inside the 
star is known, we can solve Eq.~(\ref{Peq}) and
Eqs.~(\ref{be1})-(\ref{be3}) for $P$, $h$, $f$, $A_0$ 
with a given function $G_4(X)$. 
In doing so, we need to impose regular boundary conditions 
at the center of star. 

The general relativistic (GR) solution corresponds to the vanishing 
temporal component, i.e., 
\be
{\rm GR}:~A_0=0\,.
\label{GRA0}
\ee
We also consider a nonvanishing vector-field solution (VS)
characterized by a radial-dependent temporal component $A_0(r)$ 
with the asymptotic behavior $A_0(r) \to 0$ as $r \to \infty$, i.e., 
\be
{\rm VS}:~A_0(r) \neq 0\,,\quad {\rm and} \quad
A_0(r \to \infty)=0\,.
\label{VSso}
\ee
A simple model which may allow for the existence of both 
(\ref{GRA0}) and (\ref{VSso}) is given by  
$G_{4,X}=\beta={\rm constant}$, i.e., 
\be
G_4(X)=\frac{M_{\rm pl}^2}{2}+\beta X\,,
\label{G4model}
\ee
where the first term on the right hand side corresponds 
to the Einstein-Hilbert term. 
We can also think of other couplings 
including nonlinear terms of $X$ like 
$G_4(X)=M_{\rm pl}^2/2+\sum_{n=1}\beta_n X^n$,
for instance 
$G_4(X)=(M_{\rm pl}^2/2)e^{2\beta X/M_{\rm pl}^2}$, 
but we will focus on the model (\ref{G4model}) in this paper.
Note that if the linear term of  $X$ in $G_4(X)$ is absent,
there may not be VS solutions,
as in this case the GR branch is expected to be linearly stable 
and they may be formed from other initial conditions.  

\subsection{Boundary conditions}

Let us derive general boundary conditions of NSs 
at $r=0$ and at spatial infinity for the model given by Eq.~(\ref{G4model}). 
At the center of star, we impose the regular boundary conditions 
$P'(0)=\rho'(0)=h'(0)=f'(0)=A_{0}'(0)=0$. 
Then, the solutions around $r=0$ are expressed in the forms,
\ba
& &
P(r)=p_c+\sum_{i=2}^{\infty} p_i r^i\,,\qquad 
\rho(r)=\rho_c+\sum_{i=2}^{\infty} \rho_i r^i\,,\nonumber\\
& &
h(r)=1+\sum_{i=2}^{\infty}h_i r^i\,,\qquad
f(r)=f_0+\sum_{i=2}^{\infty}f_i r^i\,,\qquad
A_0(r)=A_c+\sum_{i=2}^{\infty}\alpha_i r^i\,,
\label{expand}
\ea
where $p_c, p_i, \rho_c, \rho_i, h_i, f_0, f_i,A_c,\alpha_i$ 
are constants. 
Substituting Eq.~(\ref{expand}) into Eq.~(\ref{Peq}) and 
Eqs.~(\ref{be1})-(\ref{be3}), the iterative solutions 
around $r=0$ are given by 
\ba
P(r) &=& P_c-\frac{f_0(\rho_c+P_c)[\rho_c (f_0 M_{\rm pl}^2
+\beta A_c^2-8 \beta^2 A_c^2)+3P_c (f_0 M_{\rm pl}^2
-\beta A_c^2)]}
{12(f_0 M_{\rm pl}^2-\beta A_c^2)^2}r^2+{\cal O}(r^4)\,,
\label{so1} \\
h(r) &=& 1-\frac{f_0 \rho_c}{3(f_0 M_{\rm pl}^2-\beta A_c^2)}r^2
+{\cal O}(r^4)\,,\label{so2}\\
f(r) &=& f_0+\frac{f_0^2 [\rho_c (f_0 M_{\rm pl}^2
+\beta A_c^2-8 \beta^2 A_c^2)+3P_c (f_0 M_{\rm pl}^2
-\beta A_c^2)]}
{6(f_0 M_{\rm pl}^2-\beta A_c^2)^2}r^2+{\cal O}(r^4)\,,
\label{so3} \\
A_0(r) &=& A_c+\frac{A_c \beta f_0 \rho_c}
{3(f_0 M_{\rm pl}^2-\beta A_c^2)}r^2+{\cal O}(r^4)\,.
\label{so4}
\ea
We consider a NS with the radius $r_s$ determined by the condition 
\be
P(r_s)=0\,.
\ee
Outside the star ($r>r_s$), both $P(r)$ and $\rho(r)$ vanish. 
The boundary conditions at spatial infinity are 
\be
h(r \to \infty)=1\,,\qquad 
f(r \to \infty)=1\,,\qquad 
A_0(r \to \infty)=0\,.
\label{hfA}
\ee
Under the time reparametrization in the metric (\ref{metric}), 
the asymptotic value of $f$ can be chosen as an arbitrary constant. 
After performing the replacements $A_c=\sqrt{f_0} \bar{A}_c$, 
$A_0(r)=\sqrt{f_0} \bar{A}_0 (r)$, and $f(r)=f_0\bar{f}(r)$ 
in Eqs.~(\ref{so1})-(\ref{so4}), 
the constant $f_0$ disappears in the expressions of 
$P(r)$, $h(r)$, $\bar{f}(r)$, and $\bar{A}_0 (r)$.
In other words, we can choose $f_0=1$ without loss of generality. 

The mass function ${\cal M}(r)$ is defined by 
\be
h(r)=1-\frac{2G{\cal M}(r)}{r}\,.
\label{hM}
\ee
The Arnowitt-Deser-Misner (ADM) mass is given by 
the asymptotic value of ${\cal M}(r)$ at spatial infinity, i.e., 
\be
M \equiv \lim_{r \to \infty} {\cal M}(r)=
\frac{r}{2G} \left[ 1-h(r) \right] \biggl|_{r \to \infty}\,.
\ee
We introduce the compactness of star, as 
\be
{\cal C}=\frac{GM}{r_s}\,,
\label{compact}
\ee
where $G$ is the Newton gravitational constant.
For a given EOS, the radius and mass of a NS are known 
by numerically integrating Eqs.~(\ref{Peq}) and (\ref{be1})-(\ref{be3}) 
with the boundary conditions (\ref{so1})-(\ref{so4}) at $r=0$. 
Because of the reflection symmetry under $A_{\mu} \to -A_{\mu}$, 
we assume that $A_0 \geq 0$ in the rest of paper 
without loss of generality.
Note that we can go back to the conventional units with $G$ and $c$
by the replacement of $M_{\rm pl} \to c^2/\sqrt{8\pi G}$ and $\rho\to \rho c^2$
in the above equations.

\subsection{Equations of state of NSs}

As an EOS of relativistic stars, we first discuss the case of 
constant density $\rho$ in Sec.~\ref{consec}.
Then, in Secs.~\ref{0node} and \ref{1node}, we will 
proceed to the analysis of NS structures with a 
nonvanishing VS for two more realistic EOSs: SLy and  BSk20. 
For the latter EOSs, we introduce the dimensionless quantities, 
\ba
\xi &\equiv&\log_{10} (\rho/{\rm g \cdot cm}^{-3})=
\alpha_1+\alpha_2 \ln y \,,\\
\zeta &\equiv& \log_{10} (P/{\rm dyn \cdot cm}^{-2})
=\alpha_3+\alpha_2 \ln z\,,
\label{xizeta}
\ea
where $\alpha_1=\ln( \rho_0/{\rm g\cdot cm^{-3}})/\ln10$, 
$\alpha_2=(\ln10)^{-1}$, 
$\alpha_3=\ln( \rho_0 c^2/{\rm dyn\cdot cm^{-2}})/\ln10$, and 
\be
y \equiv \frac{\rho}{\rho_0}\,,\qquad 
z \equiv \frac{P}{\rho_0}\,.
\label{yz}
\ee
Here, $\rho_0$ is the density defined by 
\be
\rho_0 \equiv m_{\rm n} n_0
=1.6749 \times 10^{14}~{\rm g} \cdot {\rm cm}^{-3}\,,
\ee
where $m_{\rm n}=1.6749 \times 10^{-24}$~g is the neutron mass 
and $n_0=0.1~{\rm (fm)}^{-3}$ is the typical number density of NSs. 
SLy and BSk20 EOSs are parameterized as
\ba
\zeta (\xi) &=&
\frac{a_1+a_2 \xi+a_3 \xi^3}{1+a_4 \xi}
\left\{ \exp[a_5(\xi-a_6)]+1 \right\}^{-1}
+\left( a_7+a_8 \xi \right) 
\left\{ \exp[a_9(a_{10}-\xi)]+1 \right\}^{-1}
\nonumber \\
& &
+\left( a_{11}+a_{12} \xi \right) 
\left\{ \exp[a_{13}(a_{14}-\xi)]+1 \right\}^{-1}
+\left( a_{15}+a_{16} \xi \right) 
\left\{ \exp[a_{17}(a_{18}-\xi)]+1 \right\}^{-1}
\nonumber \\
& & 
+\frac{a_{19}}{1+[a_{20}(\xi-a_{21})]^2}
+\frac{a_{22}}{1+[a_{23}(\xi-a_{24})]^2}\,.
\label{zeta}
\ea
For SLy, the coefficients $a_{1, \cdots , 18}$ are given in 
Table 1 of Ref.~\cite{Haensel:2004nu}, with 
$a_{19}=a_{20}=a_{21}=a_{22}=a_{23}=a_{24}=0$. 
For BSk20, the coefficients are presented in 
Ref.~\cite{Potekhin:2013qqa} with the correspondence 
$a_i=\bar{a}_i$ for $1 \le i \le 9$, 
$a_{10}=\bar{a}_6$ and $a_i =\bar{a}_{i-1}$ for 
$11 \le i \le 24$, where $\bar{a}_i$
are the values given in the center of 
Table 2 in Ref.~\cite{Potekhin:2013qqa}.

\section{Relativistic stars with constant density}
\label{consec}

In this section, we consider relativistic stars with the constant 
density $\rho_c$ to understand properties of the nonvanishing 
VS analytically.

In the absence of the vector field $A_{\mu}$, there exist 
analytic solutions to Eqs.~(\ref{Peq}), (\ref{be1}), 
and (\ref{be2}). The metric components inside 
the star ($r \le r_s$) are given by 
\be
f=\left[ \frac{3}{2} \sqrt{1-2{\cal C}}-\frac{1}{2} 
\sqrt{1-2{\cal C} \frac{r^2}{r_s^2}} \right]^2\,, 
\qquad 
h=1-2{\cal C} \frac{r^2}{r_s^2}\,, 
\label{fin}
\ee
with 
\be
\frac{P}{\rho_c}=\frac{\sqrt{1-2{\cal C}\,r^2/r_s^2}-\sqrt{1-2{\cal C}}}
{3\sqrt{1-2{\cal C}}-\sqrt{1-2{\cal C}\, r^2/r_s^2}}\,,\qquad 
{\cal C}=\frac{\rho_c r_s^2}{6 M_{\rm pl}^2}\,.
\label{Pin}
\ee
The geometry outside the star ($r>r_s$) is described by the 
Schwarzschild metric,
\be
f=h=1-2{\cal C} \frac{r_s}{r}\,. 
\label{fout}
\ee
{}From $r=0$ to $r=r_s$, the function $f$ 
increases from $(3\sqrt{1-2{\cal C}}-1)^2/4$ to $1-2{\cal C}$. 
This increase of $f$ leads to the decrease of $P$ 
as a function of $r$ according to Eq.~(\ref{Peq}).
The stellar radius $r_s$ is determined by the point 
at which $P$ vanishes.
{}From Eqs.~(\ref{fin}) and (\ref{fout}), the function
$h$ reaches a minimum at $r=r_s$ and it starts to grow
for $r>r_s$. 

As we observe in Eqs.~(\ref{be1}) and (\ref{be2}), 
the existence of nonvanishing $A_0$ affects the metric 
components $h$ and $f$, 
so that Eqs.~(\ref{fin}) and (\ref{fout}) are 
subject to modifications. Before addressing this point, 
we first derive analytic solutions to $A_0(r)$ under some 
approximations to extract general properties of 
nonvanishing VSs. 
We then study the full numerical solution to $A_0(r)$ 
and discuss its effect on the metrics.

\subsection{Approximate vector-field solutions 
in the weak gravitational background}
\label{anaessec}

In the presence of $A_{\mu}$, the temporal vector 
component inside the star obeys
\be
A_0''+\left( \frac{2}{r}-\frac{f'}{2f}+\frac{h'}{2h} \right)
A_0'-\frac{\beta (2f \rho_c+h A_0'^2)}
{h(f M_{\rm pl}^2-\beta A_0^2)}A_0=0\,,
\label{A0eq}
\ee
which allows the existence of the GR branch (\ref{GRA0}).
To discuss whether the GR solution can be unstable to reach 
a VS with nonvanishing $A_0$, we take into account the perturbation 
$\delta A_0(r)$ around $A_0=0$.
Then, $\delta A_0(r)$ has the negative mass squared 
$m_{\rm eff}^2=2\beta \rho_c/M_{\rm pl}^2$ for 
\be
\beta<0\,,
\ee
which is at least necessary for spontaneous vectorization 
to occur.

If we consider the weak gravitational background 
with ${\cal C} \ll 1$, the following  two conditions are satisfied, 
\be
\frac{1}{r} \gg \left\{\left| \frac{f'}{f} \right|,
\left| \frac{h'}{h} \right| \right\}\,,\qquad 
\left| h-1 \right| \ll 1\,.
\label{apro1}
\ee
In realistic NSs the conditions (\ref{apro1}) can be violated, 
but we will temporally use them for the purpose of 
deriving analytic solutions approximately.
Let us also discuss the case in which $A_0'^2$ and 
$A_0^2$ are in the ranges, 
\be
f \rho_c  \gg h A_0'^2\,,\qquad 
f M_{\rm pl}^2  \gg |\beta| A_0^2\,.
\label{apro2}
\ee
Under the conditions (\ref{apro1}) and (\ref{apro2}),
Eq.~(\ref{A0eq}) is approximately given by 
\be
A_0''+\frac{2}{r} A_0'-\frac{12\beta {\cal C}}
{r_s^2} A_0 \simeq 0\,.
\label{A0eq2}
\ee
Let us derive the solution to Eq.~(\ref{A0eq2}) 
for $\beta<0$.
Imposing the boundary conditions 
$A_0(r=0)=A_c={\rm constant}$ and 
$A_0'(r=0)=0$ at the center, the resulting 
internal solution (for $r \le r_s$) is  
\be
A_0(r) \simeq A_c 
\frac{\sin (b\, r/r_s)}{b\,r/r_s}\,,
\label{A0int}
\ee
where 
\be
b \equiv \sqrt{-12\beta{\cal C}}\,.
\ee
Outside the star the density $\rho_{c}$ vanishes, while 
$hA_0'^2$ does not, so the first condition of 
Eq.~(\ref{apro2}) is violated. 
Dropping the term $hA_0'^2$ for the moment 
and employing the first condition of Eq.~(\ref{apro1}) 
outside the star, Eq.~(\ref{A0eq}) approximately 
reduces to 
\be
A_0''+\frac{2}{r} A_0' \simeq 0\,.
\label{A0eq3}
\ee
In this case, the external solution reads
\be
A_0(r) \simeq A_{\infty}+\frac{Q}{r}\,,
\label{A0inf}
\ee
where $A_{\infty}$ and $Q$ are constants.
Matching the two solutions (\ref{A0int}) and (\ref{A0inf}), and their first derivatives, respectively
at $r=r_s$, we obtain the following two relations,
\be
\frac{A_c}{A_{\infty}} \simeq \frac{1}
{\cos b}\,,\qquad 
\frac{Q}{A_c r_s} \simeq 
\frac{\sin b}{b}-\cos b\,.
\label{AQre}
\ee

For $b \simeq \pi/2-0$, there exists the 
VS with $A_{\infty} \simeq +0$ and the positive charge $Q=(2/\pi)A_c r_s$. 
{}From Eq.~(\ref{A0int}), we have $A_0(r)=(2/\pi)A_c$ 
at $r=r_s$. In this case, the temporal vector component monotonically 
decreases toward the asymptotic value $A_{\infty} \simeq 0$ without 
crossing $A_0(r)=0$. 
This VS is called the {\it 0-node} solution. 
Under the above approximation scheme, the criterion for the existence 
of 0-node solutions is that $\beta$ is smaller than the critical value $\beta_{\rm c}$ 
satisfying $b=\pi/2$, i.e., 
$\beta<-\pi^2/(48{\cal C})$.
For NSs with ${\cal C}=0.2$, this condition translates to $\beta<-1$. 

For $b \simeq 3\pi/2-0$, we also have the other VS with 
$A_{\infty} \simeq -0$ and the negative charge 
$Q=-(2/3\pi) A_c r_s$. 
In this case, the VS crosses $A_0(r)=0$ 
at $r=2r_s/3$ inside the star and reaches the
negative value 
$A_0(r)=-2A_c/(3\pi)$ at $r=r_s$. Then, $A_0(r)$ increases 
according to $A_0(r) \simeq Q/r$ with $Q<0$ toward the 
asymptotic value $-0$. 
This VS is called the {\it 1-node} solution, which crosses 
the point $A_0(r)=0$ once. 

In general, the VS with $n$ nodes corresponds to 
$b \simeq (2n+1)\pi/2-0$ with the charge 
$Q=(-1)^n(2/\pi)(2n+1)^{-1}A_c r_s$.
In this case, the vector field crosses $A_0(r)=0$ for 
$n$ times with the asymptotic behavior 
$A_0(r) \simeq Q/r$ at spatial infinity. 
For even and odd $n$, the charge $Q$ is positive 
and negative, respectively.

\subsection{Full vector-field solutions}

The relations (\ref{AQre}) have been derived by exploiting the 
conditions (\ref{apro1}) and (\ref{apro2}) to Eq.~(\ref{A0eq}). 
However, these conditions can be violated for realistic NSs with 
${\cal C}={\cal O}(0.1)$.  
If we use the Schwarzschild interior solution (\ref{fin}) to estimate the 
term in front of $A_0'$ in Eq.~(\ref{A0eq}), it follows that 
\be
\frac{2}{r}-\frac{f'}{2f}+\frac{h'}{2h} \simeq
\frac{2}{r} \left[ 1-{\cal O}(1)\,{\cal C} \frac{r^2}{r_s^2} 
\right]\,.
\label{fries}
\ee
As we will see later, the
metric functions $f$ and $h$ are subject to modifications 
by the backreaction of $A_0$.
Still, the estimation (\ref{fries}) is sufficient for the purpose of
understanding the metric corrections to the leading-order term $2/r$.
{}From Eq.~(\ref{fries}), the compactness ${\cal C}$ of order 0.1 
works to reduce the friction term in Eq.~(\ref{A0eq}), 
whose effect is particularly strong around $r=r_s$.
The metric component $h$, which appears in the denominator 
of the third term on the left hand side of Eq.~(\ref{A0eq}), 
reaches a minimum value $1-{\cal O}(1){\cal C}$ around $r=r_s$.
These properties show that, in comparison to the weak gravitational 
background with ${\cal C}$ much smaller than 0.1, the decrease of 
$A_0(r)$ inside the star is faster than that estimated by Eq.~(\ref{A0int}). 
In other words, when $\beta=-{\cal O}(1)$, the solutions can enter 
the negative $A_0(r)$ region, in which case the 0-node solution can 
disappear. Instead, it  may be possible to realize the 1-node 
solution even for $\beta=-{\cal O}(1)$.
It is also expected that the 0-node solution may be present 
for $|\beta|$ smaller than the order 1.

The field derivative $hA_0'^2$ in Eq.~(\ref{A0eq}) leads to the 
faster decrease of $A_0$ inside the star as well, whose effect is largest 
around $r=r_s$. The term $-\beta A_0^2$ gives rise to the contribution
to $fM_{\rm pl}^2$ for $A_0 >{\cal O}(0.1) M_{\rm pl}$, 
but the modification tends to be unimportant for increasing $r$ 
due to the decrease of $|A_0(r)|$. 
We recall that the  term $hA_0'^2$ is present even 
outside the star. Let us estimate the correction induced by this term 
to the solution (\ref{A0inf}). In doing so, we consider the regime $r \gg r_s$ 
in which $f$ and $h$ are close to 1 with $f M_{\rm pl}^2 \gg |\beta|A_0^2$.
Then, Eq.~(\ref{A0eq}) reduces to 
\be
A_0''+\frac{2}{r}A_0' \simeq 
\frac{\beta}{M_{\rm pl}^2}A_0'^2 A_0\,.
\label{A0ap}
\ee
Substituting the leading-order solution (\ref{A0inf}) to 
the right hand side of Eq.~(\ref{A0ap}), we obtain the 
integrated solution, 
\be
A_0(r) \simeq \left( 1+\frac{\beta Q^2}{2M_{\rm pl}^2 r^2} 
\right)A_{\infty}+\left( 1+\frac{\beta Q^2}{6M_{\rm pl}^2 r^2} 
\right) \frac{Q}{r}\,.
\label{A0ex}
\ee
The negative coupling $\beta$ works to suppress the amplitude of $A_0(r)$, 
but it still has the dependence $A_0(r) \simeq Q/r$ for sufficiently 
large $r$.
The internal and external solutions of $A_0(r)$ discussed above join
each other at $r=r_s$.

In order to confirm the presence of 0-node as well as 1-node solutions,
we numerically solve the full background equations with the 
boundary conditions (\ref{so1})-(\ref{so4}) around $r=0$. 
For this purpose, we define the dimensionless quantities, 
\be
m \equiv \frac{3{\cal M}}{4\pi \rho_0 r_0^3}\,,\qquad 
\bar{A}_0 \equiv \frac{A_0}{M_{\rm pl}}\,,\qquad
s \equiv \ln \frac{r}{r_0}\,,
\label{myzs}
\ee
where
\be 
r_0 \equiv \frac{c}{\sqrt{G\rho_0}}=89.664~{\rm km}\,.
\ee
On using Eqs.~(\ref{Peq}) and (\ref{be1})-(\ref{be3}) with Eq.~(\ref{hM}), 
we can derive the differential equations for $f$, $m$, $\bar{A}_0$, and 
$z=P/\rho_0$ with respect to the variable $s$. 
In this section, we are considering the star with constant 
$y=\rho/\rho_0$ for $r \le r_s$.

We find that the 0-node solution is present for 
\be
-2 \lesssim \beta \lesssim -0.1\,.
\ee
Thus, there exists the 
0-node solution even for $\beta=-{\cal O}(0.1)$. 
The numerical computation also shows the existence 
of 1-node solutions for 
\be
\beta \lesssim -2\,.
\ee
As we discussed above,  strong gravitational effects 
associated with $f, h$ and the existence of term $hA_0'^2$ in Eq.~(\ref{A0eq})
lead to the larger effective coupling $\beta$, so that the nonvanishing 
VS tends to 
enter the negative $A_0$ region for $\beta=-{\cal O}(1)$.
For $\beta \lesssim -{\cal O}(10)$, we also find the existence of 
2-node solutions (and plausibly higher-node solutions),
but we will not consider the regime of such 
large coupling $|\beta|$.

Note that
our model corresponds to the special case of 
$\eta=-2\Omega$ in the notation of Ref.~\cite{Annulli},
and the correspondence with our notation is $\beta=\Omega/2$.
Equation (23) in Ref.~\cite{Annulli} shows that 
under the weak gravitational approximation
the tachyonic instability of a GR constant density star in the polar 
perturbation sector would appear below $\beta\approx -\pi^2/(12{\cal C})$,
which would be of ${\cal O}(-5)$
when extrapolated to the compactness ${\cal C}= {\cal O}(0.15)$.
Thus, the value of $\beta$ for the 1-node solution
is somewhat consistent with that for the critical coupling 
associated with the onset of tachyonic instability.
This motivates us to study 
the 1-node VS in more details.

\begin{figure}[h]
\begin{center}
\includegraphics[height=3.5in,width=3.3in]{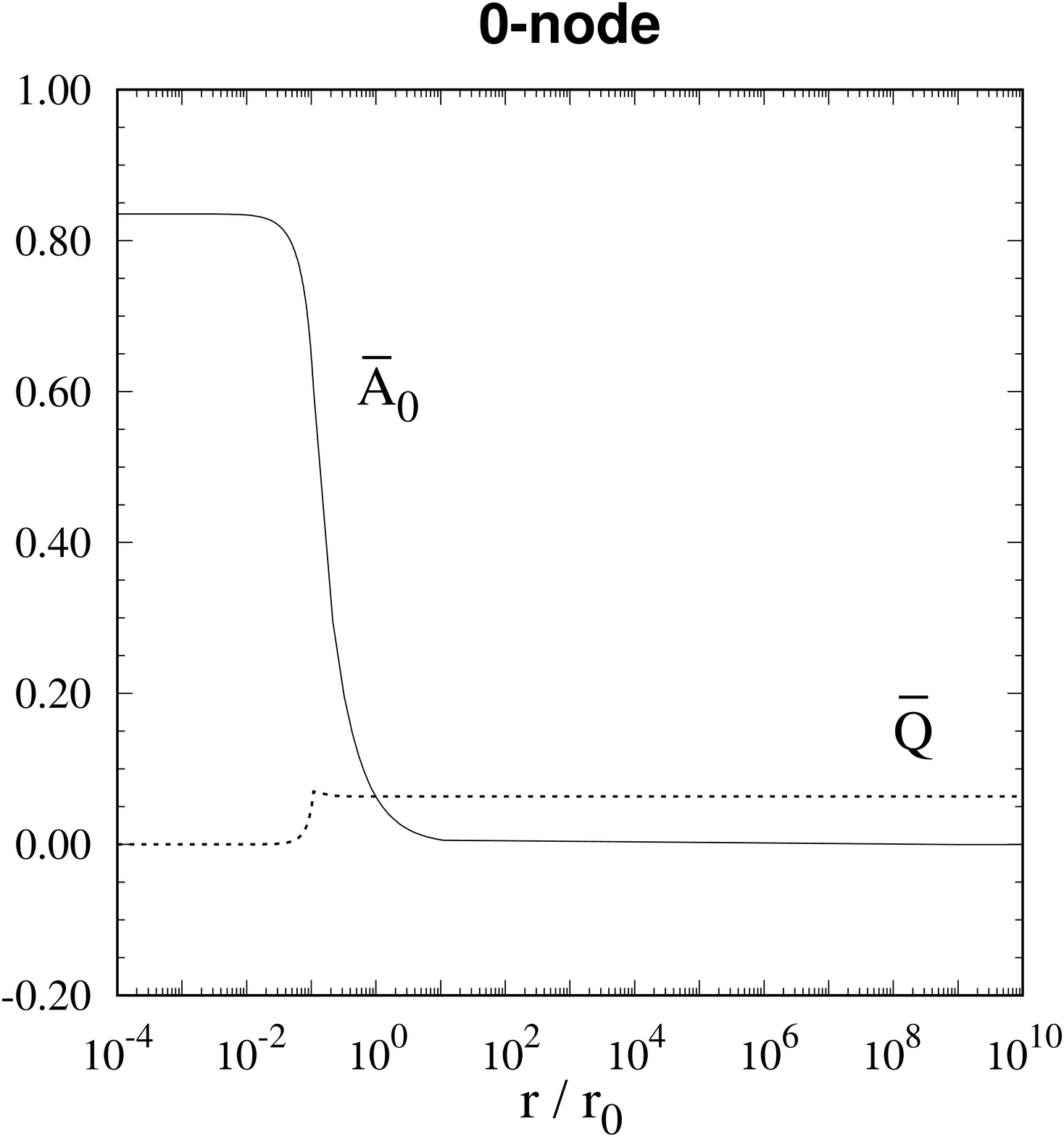}
\includegraphics[height=3.5in,width=3.3in]{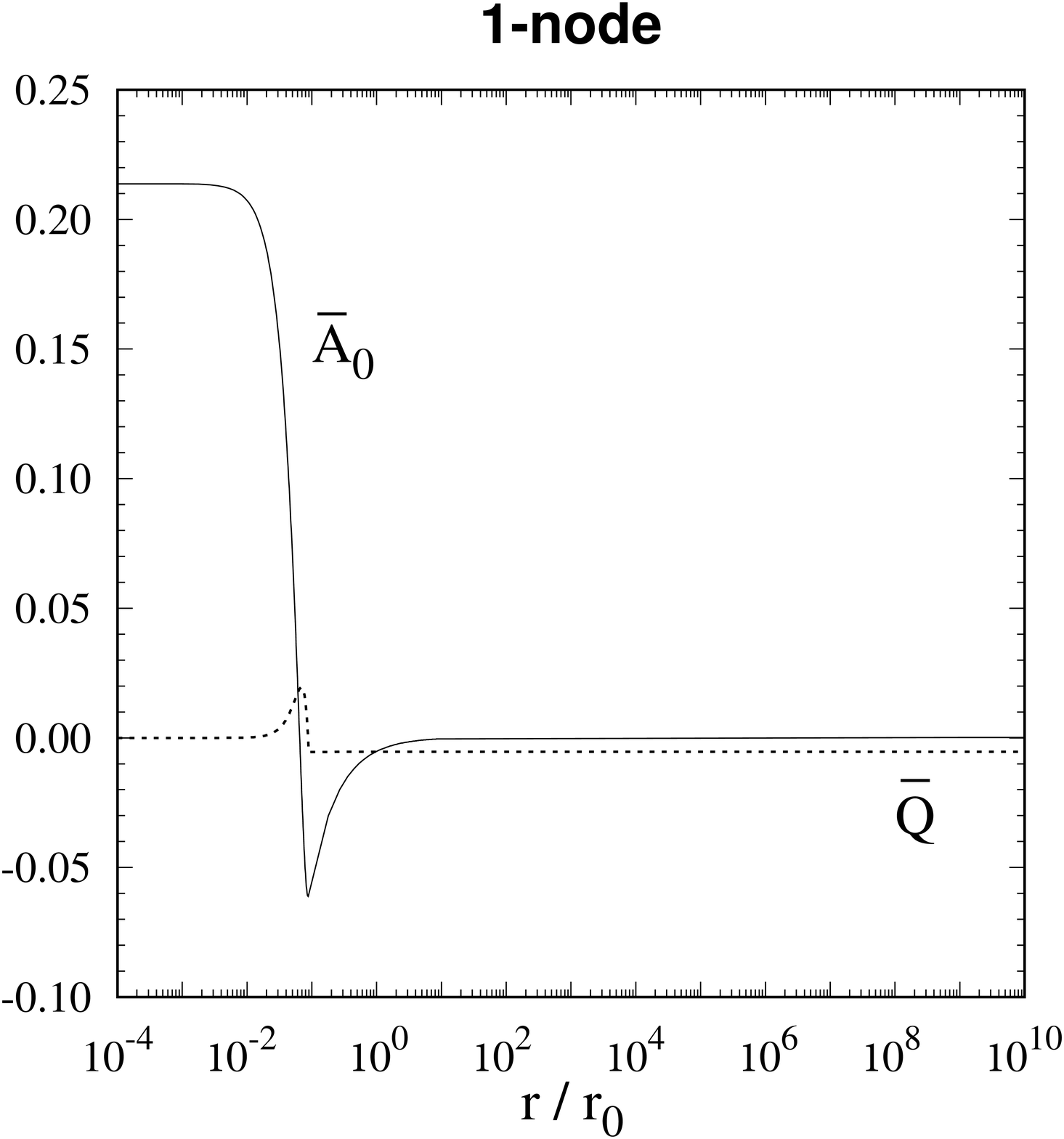}
\end{center}
\caption{\label{fig1} 
(Left) Example of the 0-node solution showing 
$\bar{A}_0=A_0/M_{\rm pl}$ and $\bar{Q}=-r^2 A_0'/(r_0 M_{\rm pl})$ 
versus $r/r_0$
for $\beta=-0.4$, $A_c=0.83525M_{\rm pl}$, $\rho_c=7\rho_0$, 
and $P_c=0.1946\rho_0$. 
(Right) Example of the 1-node solution showing
$\bar{A}_0=A_0/M_{\rm pl}$ and 
$\bar{Q}=-r^2 A_0'/(r_0 M_{\rm pl})$ versus $r/r_0$
for $\beta=-5.0$, $A_c=0.21377 M_{\rm pl}$, 
$\rho_c=9\rho_0$, and $P_c=0.5640\rho_0$.
}
\end{figure}

In the left panel of Fig.~\ref{fig1}, we plot $\bar{A}_0$ versus $r/r_0$
corresponding to the 0-node solution for $\beta=-0.4$.
In this case, the temporal vector component continuously decreases 
with the increase of $r$ toward the asymptotic value $A_0 \simeq +0$. 
We also compute the following dimensionless quantity,
\be
\bar{Q} \equiv -\frac{r^2A_0'}{r_0 M_{\rm pl}}\,.
\label{barQ}
\ee
Provided that $A_0$ behaves as Eq.~(\ref{A0ex}) for 
$r \gg r_s$, $\bar{Q}$ should approach constant 
value $Q/(r_0 M_{\rm pl})$. 
Indeed, as we see in the left panel of Fig.~\ref{fig1}, the numerical
value of $\bar{Q}$ approaches a positive constant.
This means that the 0-node solution has a positive charge $Q$ with 
$A_0(r)$ decreasing as $\propto Q/r$ at spatial infinity.

The right panel of Fig.~\ref{fig1} shows $\bar{A}_0$ and $\bar{Q}$ 
versus $r/r_0$ for $\beta=-5$, which corresponds to the 1-node solution.
In this case, the stellar radius is $r_s=0.09r_0$, around which $A_0(r)$ 
reaches a negative minimum. Outside the star, $A_0(r)$ asymptotically 
approaches the value $-0$ with the dependence $A_0(r) \simeq Q/r$. 
Indeed, the quantity (\ref{barQ}) approaches a negative constant 
and hence the 1-node solution has a negative charge $Q$. 

In the following, we estimate the backreaction of 
vector field on the metric components $f$ and $h$. 
We recall that the iterative solutions to $P$, $h$, $f$ around $r=0$ 
are given by Eqs.~(\ref{so1})-(\ref{so3}). 
As long as $f_0 M_{\rm pl}^2 \gg |\beta| A_c^2$, 
we have $h \simeq 1-\rho_c r^2/(3M_{\rm pl}^2)$ from 
Eq.~(\ref{so2}) and hence $h$ behaves in the same manner 
as in GR.
On the other hand, the $r$ derivative of Eq.~(\ref{so3}) is given by 
\be
f'(r) \simeq \frac{f_0^2 [\rho_c (f_0 M_{\rm pl}^2
+\beta A_c^2-8 \beta^2 A_c^2)+3P_c (f_0 M_{\rm pl}^2
-\beta A_c^2)]}
{3(f_0 M_{\rm pl}^2-\beta A_c^2)^2}r\,.
\ee
The term $-8\beta^2 A_c^2$ is $-8\beta$ times as large as 
the term $\beta A_c^2$.
This means that, for $\beta<-{\cal O}(0.1)$, the former 
cannot be neglected relative to the latter. 
Moreover, both $-8\beta^2 A_c^2$ and $\beta A_c^2$ 
work to reduce the derivative $f'(r)$.
{}From Eq.~(\ref{Peq}), this means that the pressure $P(r)$ 
changes slowly toward the surface of star. 
In the pressure (\ref{so1}) there are also the terms 
$-8\beta^2 A_c^2$ and $\beta A_c^2$, which prevent the 
decrease of $P(r)$ induced by the term $f_0 M_{\rm pl}^2$.
In other words, the vector field acts as the repulsive force 
to gravity. The Schwarzschild internal solutions to $f$ and 
$P$, which are given in Eqs.~(\ref{fin}) and (\ref{Pin}), 
are no longer valid for the coupling $\beta<-{\cal O}(0.1)$.

The slow change of $P(r)$ for $\beta<-{\cal O}(0.1)$ leads to the 
radius $r_s$ and compactness ${\cal C}=\rho_c r_s^2/(6M_{\rm pl}^2)$
larger than those in GR. 
The left panel of Fig.~\ref{fig1} corresponds to the 0-node solution 
with EOS $P/\rho_c=2.78 \times 10^{-2}$ at $r=0$.
On using the GR solution (\ref{Pin}), this translates to the 
compactness ${\cal C}=0.05$. 
Solving the full background equations of motion, 
however, the actual value of compactness is found to be 
${\cal C}=0.29$. For the 0-node solution in Fig.~\ref{fig1}, 
the pressure stays nearly constant up to the distance 
$r \simeq 0.06 r_0$ due to the smallness of $f'(r)$ 
induced by the coupling $\beta$. This is followed by 
the decrease of $P(r)$ up to the surface $r_s=0.1r_0$.
This value is much larger than the corresponding radius 
$r_s=0.04r_0$ in GR. 
Thus the nonvanishing 0-node solution can be distinguished 
from the GR solution in terms of $r_s$ and ${\cal C}$.

The 1-node solution shown in the right panel of Fig.~\ref{fig1} leads to 
similar increases of $r_s$ and ${\cal C}$ relative to those in GR. 
In this case the term $-8 \beta^2 A_{c}^2$ overwhelms $f_0 M_{\rm pl}^2$ 
up to the distance $r=0.05r_0$, so the function $f$ decreases from $r=0$ up to 
this distance. This results in the growth of $P(r)$ for $0<r<0.05r_0$.
After the term $-8\beta^2 A_0^2$ becomes subdominant to 
$f_0 M_{\rm pl}^2$ with the decrease of $A_0$, the pressure starts to 
decrease toward the surface ($r_s=0.09r_0$). 
As we will see in Sec.~\ref{1node} for more realistic EOSs,  
there are cases in which $f'/f$ remains positive for the 1-node solution. 
In such cases the pressure $P(r)$ decreases outwards, but the decreasing 
rate is smaller than that in GR and hence the radius $r_s$ is larger.
 
The above discussion shows the importance of vector-field coupling 
on the metric component $f$, which in turns affects the radial 
dependence of pressure inside the star.
Depending on the coupling $\beta$ and central density $\rho_c$, 
the values of $A_c$ allowing for the asymptotic behavior 
$A_0 (r\to  \infty)=0$ are different. 
For decreasing $A_c$, both $r_s$ and ${\cal C}$ tend to approach 
those in GR.

\section{0-node NS solutions}
\label{0node}

We study the existence and property of 0-node solutions 
for realistic EOSs of NSs:~SLy and BSk20. 
In Eq.~(\ref{yz}), we introduced the dimensionless quantities 
$y$ and $z$ associated with $\rho$ and $P$, respectively.
The derivative of $y$ with respect to $s=\ln (r/r_0)$, which 
is denoted as $y_{,s}=\rd y/\rd s$, is given by 
\be
y_{,s} = \frac{y}{z} \left( \frac{{\rm d} \zeta}
{{\rm d} \xi} \right)^{-1} z_{,s} 
= -\frac{y(y+z)}{2z} \left( \frac{{\rm d} \zeta}
{{\rm d} \xi} \right)^{-1}
\frac{f_{,s}}{f}\,,
\label{dif1}
\ee
where we used Eq.~(\ref{Peq}) in the second equality. 
The dimensionless quantities $f$, $m=3{\cal M}/(4\pi \rho_0 r_0^3)$, 
$\bA_0=A_0/M_{\rm pl}$ obey the 
differential equations, 
\ba
\frac{f_{,s}}{f} &=& \frac{2f(1-h)+16\pi f e^{2s}z+2\beta
[\bA_0^2-h\bA_0 (\bA_0+4\bA_{0,s})]-h\bA_{0,s}^2}
{2h (f-\beta \bA_0^2)}\,,
\label{dif2}\\
m_{,s} &=& \frac{3e^s (16\pi f e^{2s}y+h\bA_{0,s}^2)}
{16\pi (f-\beta \bA_0^2)}\,,
\label{ms}\\
\bA_{0,ss} &=& \frac{\bA_{0,s} [4\pi f (y+z)e^{2s}-fh
+\beta \bA_0^2-\beta h \bA_0 \bA_{0,s}]
+16\pi f \beta ye^{2s} \bA_0}
{h (f-\beta \bA_0^2)}\,,
\label{dif4}
\ea
where 
\be
h=1-\frac{8\pi m}{3e^s}\,,\qquad 
z=\exp \left[ \frac{\zeta(\xi)-\alpha_3}{\alpha_2} \right]\,,
\label{hx}
\ee
with $\xi=\alpha_1+\alpha_2 \ln y$.
For EOSs (\ref{zeta}), we solve the differential Eqs.~(\ref{dif1})-(\ref{dif4})
for $y$, $f$, $m$, and $\bar{A}_0$ under the boundary conditions 
(\ref{so1})-(\ref{so4}) around $r=0$. 
For a given negative coupling $\beta$ and central density $y_c=\rho_c/\rho_0$, 
we search for the value of $A_c$ at $r=0$ approaching 
$A_0(r) \simeq 0$ for $r \gg r_s$. 
Numerically, the integration is performed up to the 
distance $r=10^{12}r_0$.

\begin{figure}[h]
\begin{center}
\includegraphics[height=3.2in,width=3.4in]{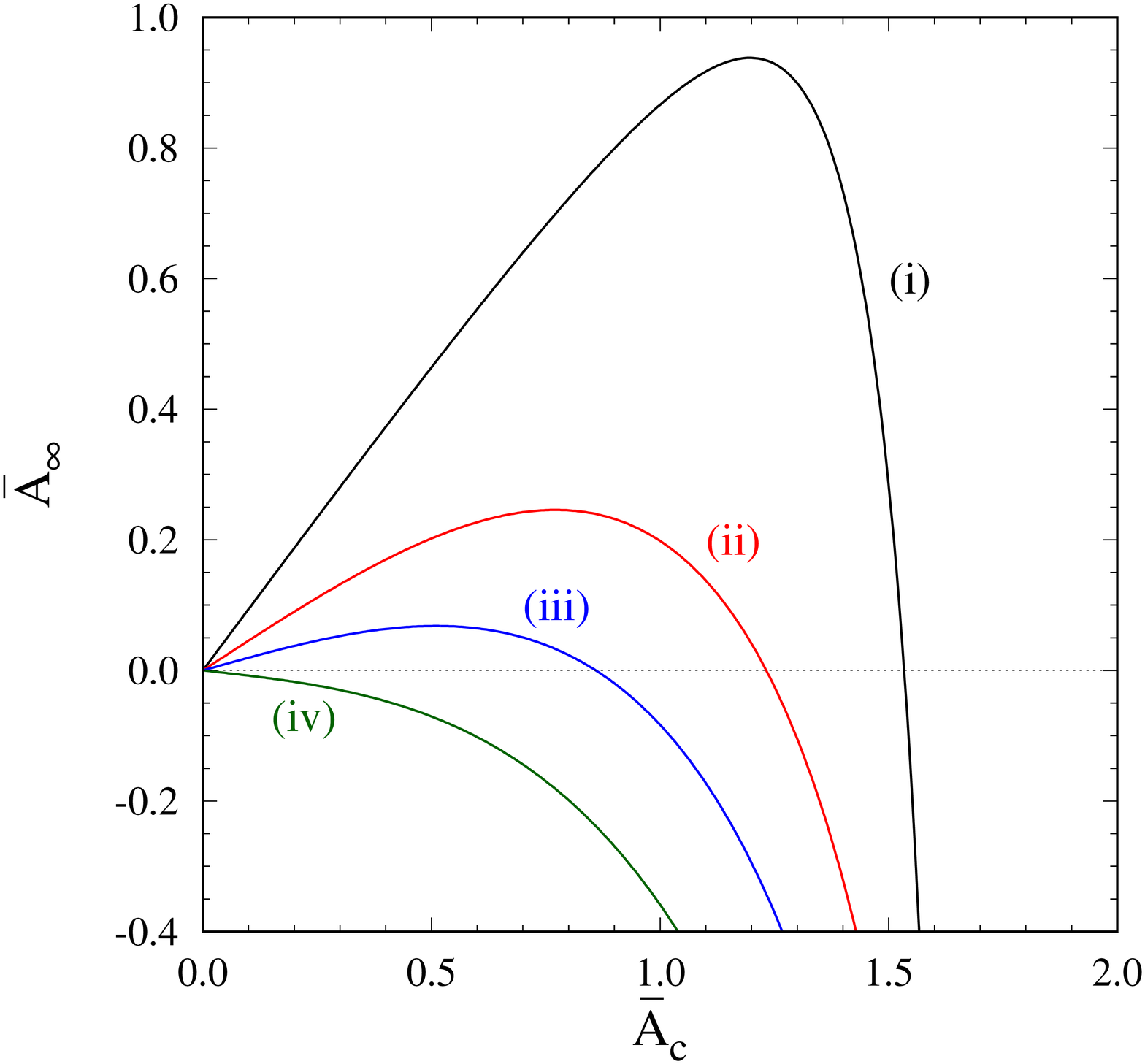}
\includegraphics[height=3.2in,width=3.4in]{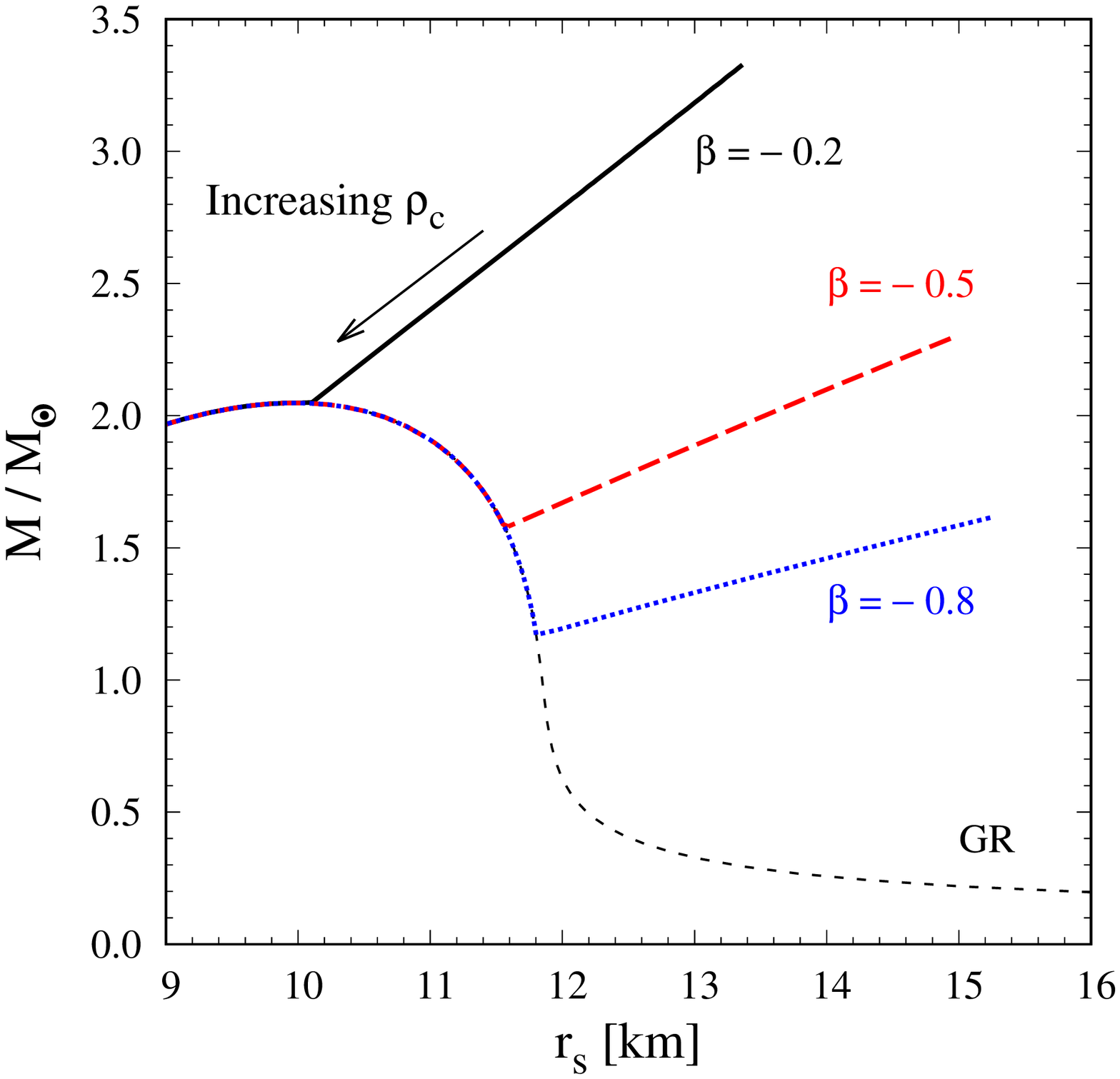}
\end{center}
\caption{\label{fig2} 
(Left) 
The temporal vector component $\bA_{\infty}$ at $r \gg r_s$ 
versus its central value $\bA_c$ for SLy EOS with $\beta=-0.2$. 
Each plot corresponds to the central densities 
(i) $\rho_c=2\rho_0$, (ii) $\rho_c=8\rho_0$, 
(iii) $\rho_c=12\rho_0$, and (iv) $\rho_c=18\rho_0$, 
respectively. The dashed line represents $\bA_{\infty}=0$.
(Right)
The ADM mass $M$ (in the unit of solar mass $M_{\odot}$) 
versus the radius $r_s$ for SLy EOS 
with three different values of $\beta$. 
For $\beta=-0.2, -0.5, -0.8$, we consider the regimes of 
central densities $\rho_c \ge 8 \rho_0$, $\rho_c \ge 4 \rho_0$, and 
$\rho_c \ge 3\rho_0$, respectively. 
We also show how the  values of $M$ and $r_s$  
shift with the increase of $\rho_c$. 
The $M$-$r_s$ relation in GR is represented 
as the thin black dashed line.
}
\end{figure}

In the left panel of Fig.~\ref{fig2}, we plot the asymptotic 
value $\bA_{\infty}=A_{\infty}/M_{\rm pl}$ at $r \gg r_s$
versus $\bar{A}_{c}=A_{c}/M_{\rm pl}$ at $r=0$ 
for SLy EOS with $\beta=-0.2$. 
The curve (i), which corresponds to the central 
density $\rho_c=2\rho_0$, has two intersections with the line 
$\bA_{\infty}=0$, i.e., $\bA_c=0$ and $\bA_c=1.53$.
The former is the GR branch, while the latter is the 0-node solution.
In Fig.~\ref{fig2}, we observe that the positive change of $\bA_c$ from 
the GR branch leads to the positive shift of $\bA_{\infty}$. 
On the other hand, the positive change of $\bA_c$ from 
the 0-node solution results in the negative shift of $\bA_{\infty}$. 
In Fig.~6 of Ref.~\cite{Annulli}, the similar behavior was found for polytrope EOS in different vector-tensor theories. 
In such cases the authors of Ref.~\cite{Annulli} showed that 
the GR solution is stable by considering axial and polar parity perturbations, 
so it is unlikely that the 0-node solution arises 
as the consequence of tachyonic instabilities of the GR solution. 
Rather, the 0-node solution discovered in our GP theories may be induced 
by some nonlinear effects such as the selected choice of initial conditions.

The values of $\bA_c$ corresponding to the 0-node branch get 
smaller for larger $\rho_c$, see cases (ii) and (iii) in the left panel 
of Fig.~\ref{fig2}.
For $\beta=-0.2$, this nonvanishing $A_0$ solution disappears above 
the critical density $\bar{\rho}_c \simeq 16\rho_0$, in which regime
only the GR branch is left. 
The case (iv) in Fig.~\ref{fig2} corresponds to such an example.
For the coupling $\beta$ of order $-0.1$, we find that the 0-node solution 
is present for both SLy and BSk20 EOSs and that the $(\bA_{\infty}, \bA_c)$ relation is similar to that shown in Fig.~\ref{fig2}. 
Below a critical central density $\bar{\rho}_c$ (which depends on 
the coupling $\beta$), there exist both GR and 0-node branches,
with theoretical curves in the  $(\bA_{\infty}, \bA_c)$ plane convex upward.  

This property is very different from the 0-node solution 
in scalar-tensor theories.
As we will briefly review in Appendix, 
scalarized solutions in scalar-tensor theories 
do not appear below a critical central density $\rho_{c1}$.
For $\rho_c<\rho_{c1}$, the field value $\phi_{\infty}$ 
at $r \gg r_s$ monotonically increases with the growth of central value 
$\phi_c$ from the GR point $(\phi_{\infty},\phi_c)=(0,0)$. 
In this regime, the GR solution is stable.
Above the critical density $\rho_{c1}$, the theoretical curves in the 
($\phi_{\infty}, \phi_c)$ plane, which are convex downward, enter the region 
$\phi_{\infty}<0$, so that the 0-node solution appears besides the GR branch. 
In the left panel of Fig.~\ref{figap} in Appendix, this transition can be seen 
from the curve (i) to (ii), where the latter has two intersecting points 
at $\phi_{\infty}=0$. 
In case (ii), the GR branch $\phi_c=0$ can be unstable to undergo 
spontaneous scalarization to the 0-node solution with $\phi_c \neq 0$.
Above a second critical density $\rho_{c2}$, the theoretical curves in the 
($\phi_{\infty}, \phi_c)$ plane again enter the region $\phi_{\infty}>0$ 
and hence there is only the GR solution in this regime (see case (v) 
in Fig.~\ref{figap}).
The 0-node solution in scalar-tensor theories, which can arise out of 
spontaneous scalarization from the GR branch, exists for 
$\rho_{c1}<\rho_c<\rho_{c2}$. 
In the numerical simulation of Fig.~\ref{figap} in Appendix,  
$\rho_{c1}=4.3\rho_0$ and $\rho_{c2}=14.4\rho_0$.

In GP theories, the fact that the convex-upward curves 
in the $(\bA_{\infty}, \bA_c)$ plane 
have the intersection point $\bA_c>0$ with $\bA_{\infty}=0$ is 
related to the property that $A_0(r)$ always decreases 
as a function of $r$ around $r=0$, see Eq.~(\ref{so4}).
Moreover, the term $-8\beta^2 A_c^2$ in Eq.~(\ref{so1}) 
always works to slow down the decrease of $P(r)$. 
This is attributed to the slower increase of $f'/f$ 
as compared to the GR branch.
These properties are different in scalar-tensor theories, in that 
the behavior of $\phi(r)$ and $P(r)$ depends on whether EOS is in the range $\rho_c>3P_c$ or not (see Appendix). 
For $\rho_c<3P_c$, the scalar field $\phi(r)$ increases 
as a function of $r$ deep inside the star, 
but this is not the case for $A_0(r)$ in GP theories. 
As we see in the left panel of Fig.~\ref{fig2}, 
the nonvanishing vector-field configuration with $A_c$ 
of order $M_{\rm pl}$ is present for a wide range 
of $\rho_c$ below the critical value 
$\bar{\rho}_c \simeq 16\rho_0$.

In the right panel of Fig.~\ref{fig2}, we show the mass $M$ versus 
the radius $r_s$ for three different values of $\beta$.
When $\beta=-0.2$, the range of $\rho_c$ plotted in the figure 
is $\rho_c \ge 8 \rho_0$. 
In this range of $\rho_c$ the derivative $f'/f$ is positive inside the star, 
so that the pressure $P(r)$ decreases outwards.
Even for $\rho_c=8 \rho_0$, we have $M=3.33M_{\odot}$ 
and $r_s=13.4$ km, both of 
which are larger than the corresponding values $M=1.74M_{\odot}$ 
and $r_s=11.4$ km in GR. These changes are mostly attributed to 
the fact that the slow decrease of $P(r)$ induced by the 
coupling $\beta$ leads to larger $r_s$. 
If we consider $\rho_c$ lower than $8\rho_0$, 
the quantity $f'/f$ around $r=0$ further gets smaller and 
hence both $M$ and $r_s$ are greater than those 
for $\rho_c=8 \rho_0$.
As $\rho_c$ increases in the region 
$\rho_c \ge 8 \rho_0$, $M$ and $r_s$ decrease and they 
finally approach those in GR. 
This is consistent with the fact that the value of $A_c$
for the 0-node solution decreases for increasing $\rho_c$ 
in the left panel of Fig.~\ref{fig2}
and only the GR branch is left for $\rho_c>16\rho_0$. 

As the coupling $|\beta|$ increases, the range of $\rho_c$
allowing for the nonvanishing VS is limited to the region with low densities. 
For $\beta=-0.5$ and $\beta=-0.8$, 
the 0-node solution
with $M$ and $r_s$ different from those in GR exists for 
$\rho_c \lesssim 7 \rho_0$ and $\rho_c \lesssim 5\rho_0$, respectively.
In Fig.~\ref{fig2} we can confirm that, for larger $|\beta|$ 
(of order 0.1), the deviation from the GR values of $M$ and 
$r_s$ in high-density regions tends to be smaller.
This behavior is also consistent with the results found in 
Ref.~\cite{Annulli} in different vector-tensor theories.
For $|\beta|={\cal O}(1)$ the 0-node solution tends to disappear, but 
the 1-node solution starts to appear as we already 
discussed in Sec.~\ref{consec}.
The numerical results shown in Fig.~\ref{fig2} are obtained 
for SLy EOS, but we confirmed that the similar property 
also holds for BSk20 EOS
and hence the results are insensitive to the choice of EOSs.

In order to check whether the 0-node solution discussed above is 
gravitationally bound, we compute the proper mass,
\be
M_{p}=\int_0^{r_s} \frac{4\pi \rho r^2}{\sqrt{h}}
{\rm d}r\,.
\ee
The gravitational binding energy is given by $\Delta=M_p-M$. 
For the 0-node branch corresponding to the numerical simulation 
of Fig.~\ref{fig2}, we find that $\Delta$ is generally  positive.  
Then, the necessary condition for gravitational stability is at least satisfied.  
However, we need to consider the axial and polar perturbations 
to judge the stability of solutions properly. 
In addition, it is not yet clear whether the 0-node solutions 
constructed above can be realized from certain initial data or not. 
It is beyond the scope of our paper to address these issues.

\section{1-node NS solutions}
\label{1node}

\begin{figure}[h]
\begin{center}
\includegraphics[height=3.1in,width=3.3in]{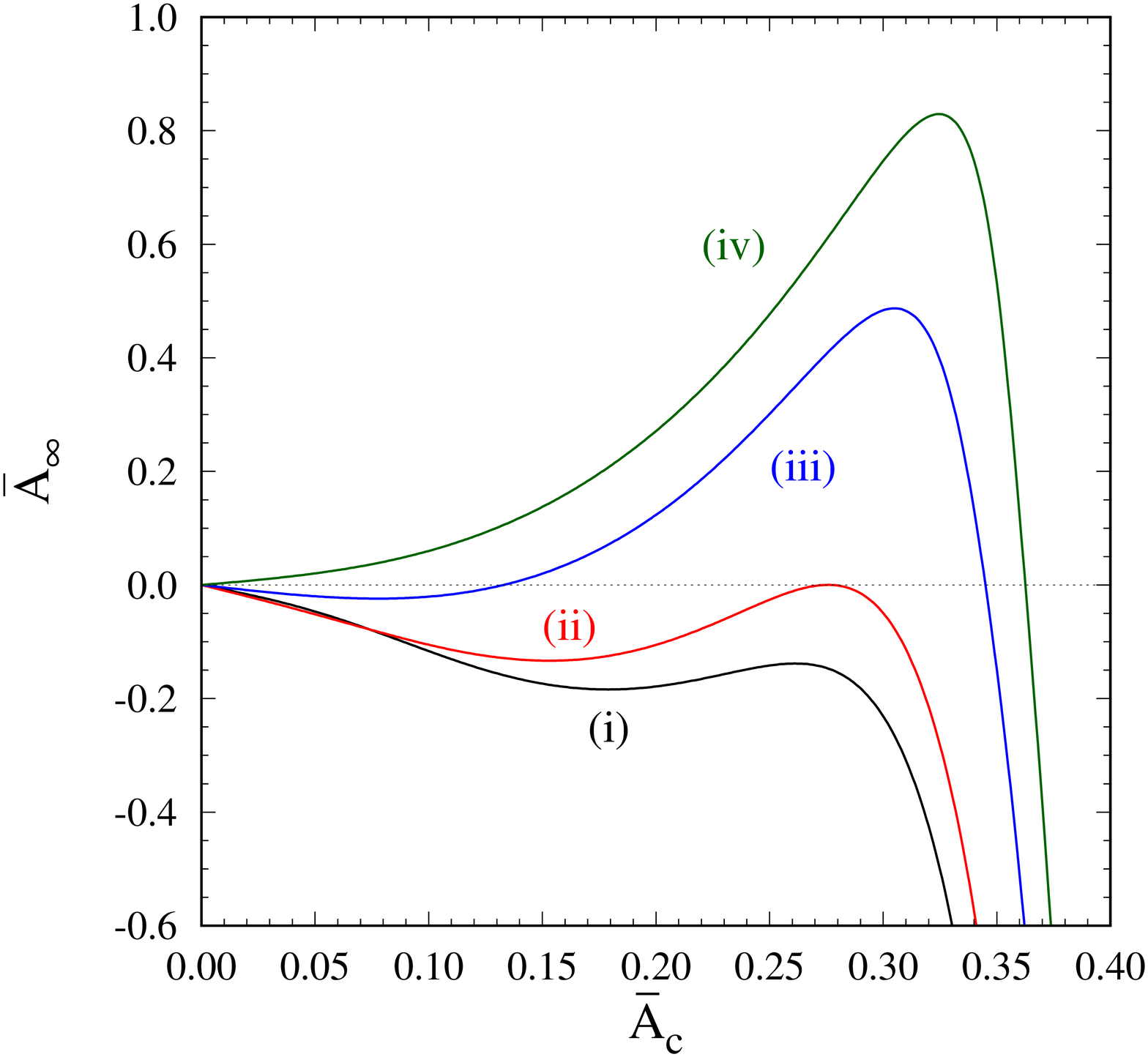}
\includegraphics[height=3.1in,width=3.3in]{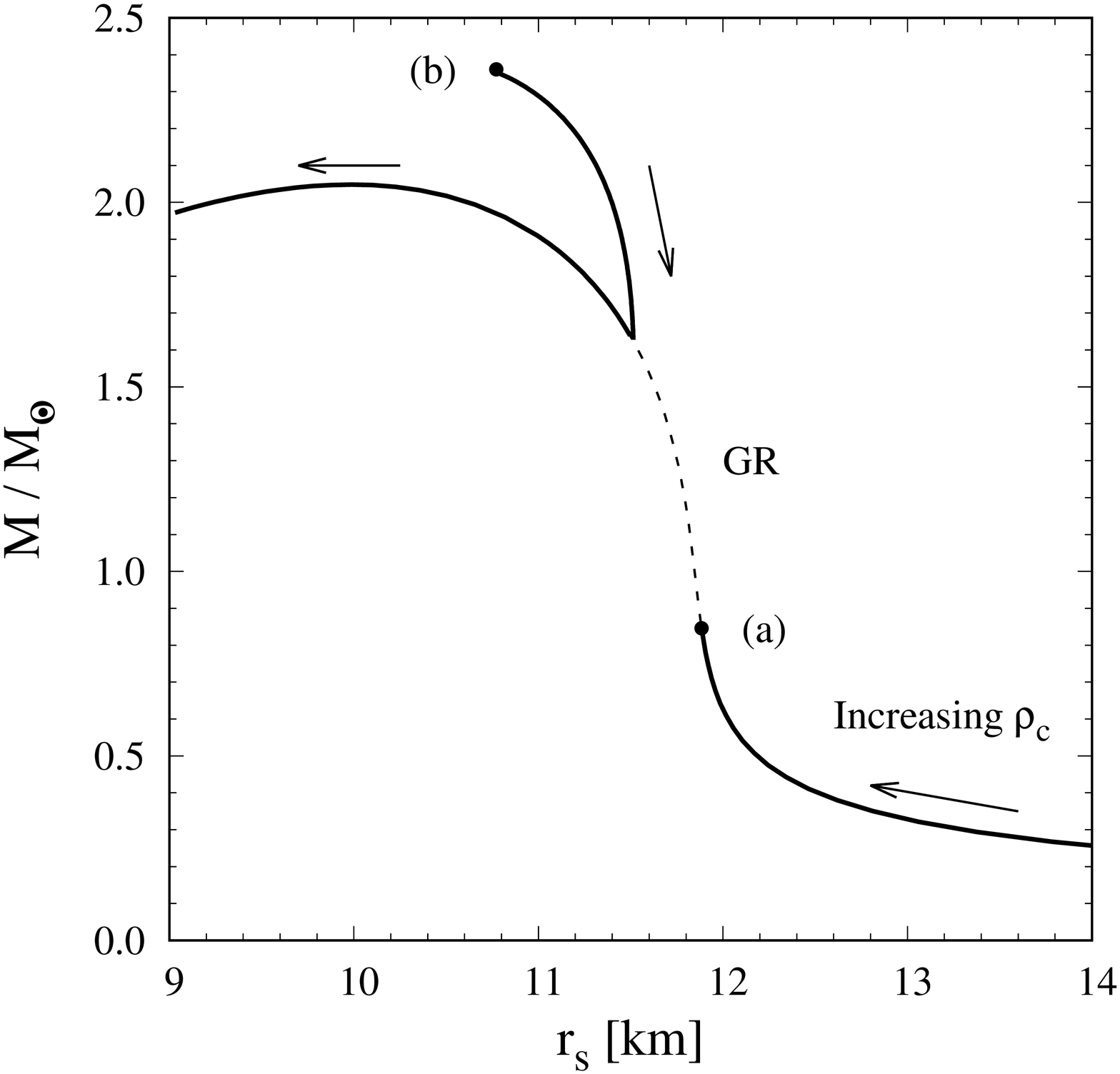}
\end{center}
\caption{\label{fig3} 
(Left) 
$\bA_{\infty}$ versus $\bA_c$ for 
SLy EOS with $\beta=-5$. 
Each plot corresponds to the central densities 
(i) $\rho_c=3\rho_0$, (ii) $\rho_c=3.74\rho_0$, 
(iii) $\rho_c=6\rho_0$, and (iv) $\rho_c=8\rho_0$, 
respectively. The dashed line represents $\bA_{\infty}=0$.
(Right)
$M/M_{\odot}$ versus the radius $r_s$ for 
SLy EOS with $\beta=-5$. 
When $\rho_c$ exceeds the critical value 
$\hat{\rho}_{c1}=3.74 \rho_0$, the 1-node VS  appears 
as the intersection point of theoretical curves in the 
$(\bA_{\infty}, \bA_c)$ plane with $\bA_{\infty}=0$, 
i.e., case (ii) in the left panel. 
At $\rho_c=\hat{\rho}_{c1}$, the GR point (a) 
jumps to the other point (b) in the
($M$, $r_s$) plane. As $\rho_c$ increases further, 
the nonzero $\bA_c$ corresponding to the smaller intersection 
value with $\bA_{\infty}=0$ tends to decrease toward $\bA_c=0$. 
Above the critical density $\hat{\rho}_{c2}=7.1 \rho_0$, 
$M$ and $r_s$ are identical to those in GR.
}
\end{figure}

\begin{figure}[h]
\begin{center}
\includegraphics[height=3.2in,width=3.3in]{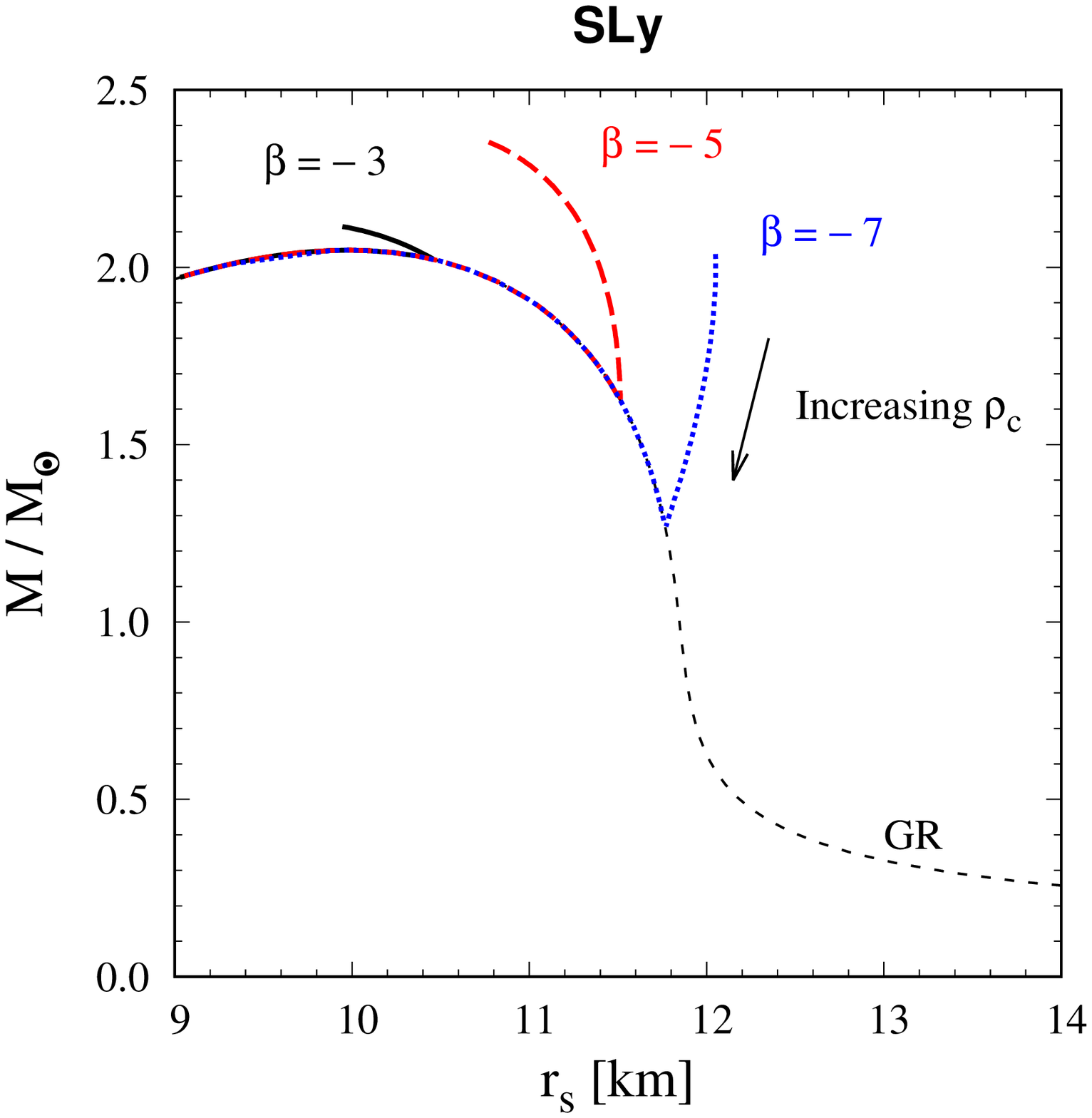}
\includegraphics[height=3.2in,width=3.4in]{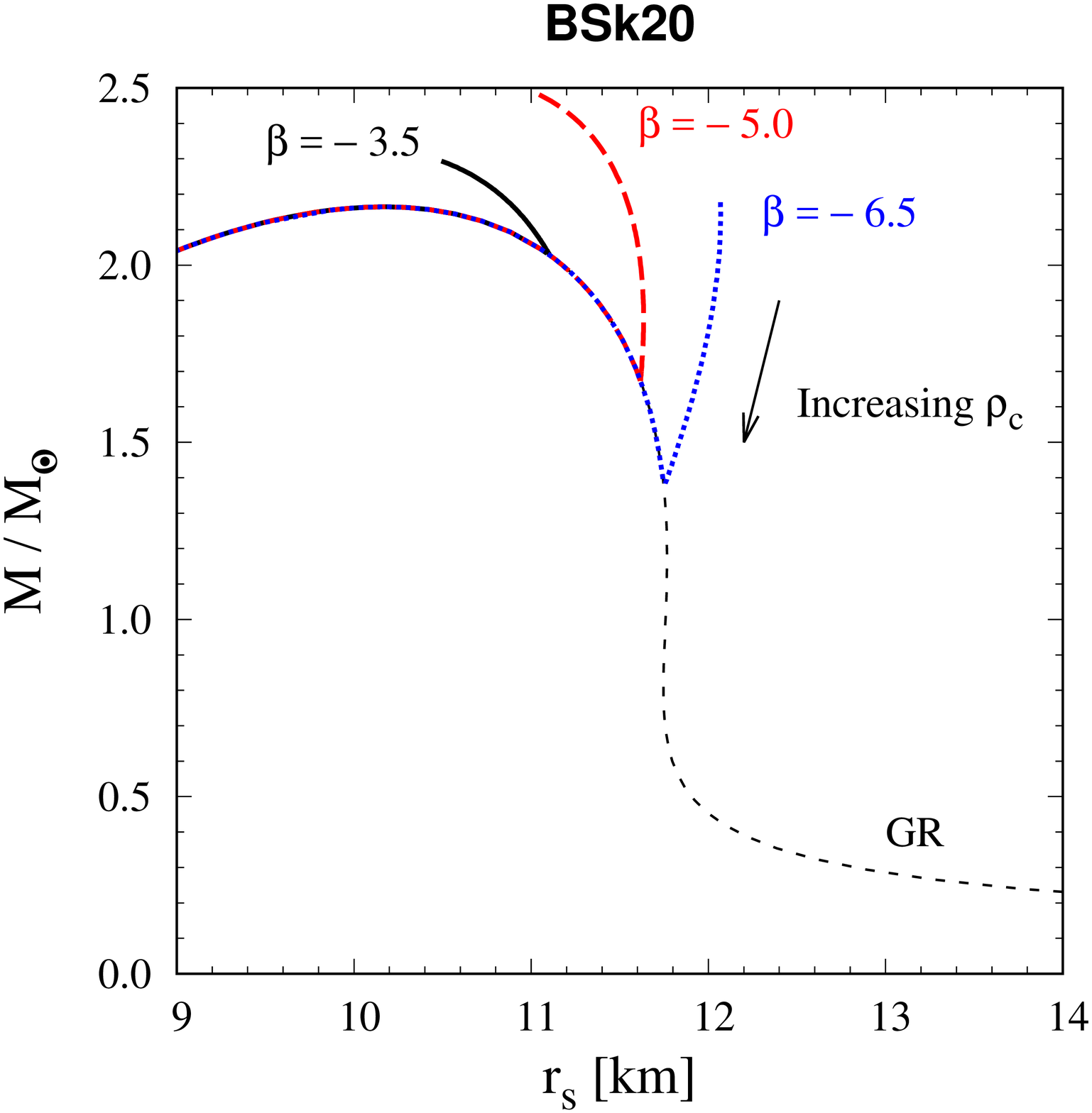}
\end{center}
\caption{\label{fig4} 
(Left) 
$M/M_{\odot}$ versus the radius $r_s$ for 
SLy EOS 
with several different values of $\beta$. 
For $\beta=-3, -5, -7$, the ranges of $\rho_c$ plotted in 
the figure are $\rho_c>12.32 \rho_0$, 
$\rho_c>3.74 \rho_0$, and $\rho_c>1.5\rho_0$, respectively.
(Right) 
$M/M_{\odot}$ versus the radius $r_s$ for  
BSk20 EOS. 
For $\beta=-3.5, -5.0, -6.5$, we consider the density regions 
$\rho_c>8.15 \rho_0$, $\rho_c>3.41 \rho_0$, 
and $\rho_c>1.66 \rho_0$, respectively.
}
\end{figure}

We then proceed to the investigation of 1-node solutions 
for SLy and BSk20 EOSs. 
As a function of $r$, the temporal vector component 
of 1-node solutions crosses 
$A_0(r)=0$ once at a finite radius 
and approaches $A_0(r) \to 0$ at spatial infinity. 
As in the case of constant density $\rho$ discussed in 
Sec.~\ref{consec}, our numerical computation shows that 
the 1-node solution exists for $\beta=-{\cal O}(1)$.

In the left panel of Fig.~\ref{fig3}, we plot the asymptotic value
$\bA_{\infty}=A_{\infty}/M_{\rm pl}$ for $r \gg r_s$ 
versus $\bA_c=A_c/M_{\rm pl}$ at $r=0$ for 
SLy EOS
with $\beta=-5$ by choosing several different central 
densities $\rho_c$. 
In case (i), which corresponds to $\rho_c=3\rho_0$, 
there is only the GR branch characterized by 
$\bA_{\infty}=0$ and $\bA_c=0$.
As $\rho_c$ increases, the intersection with 
the line $\bA_{\infty}=0$ starts to appear at 
$\bA_c>0$ for $\rho_c$ exceeding the critical 
density $\hat{\rho}_{c1}=3.74 \rho_0$.
The nonvanishing VS for $\rho_c=\hat{\rho}_{c1}$ has the 
value $\bA_c=0.2741$, see case (ii) in Fig.~\ref{fig3}. 
In the right panel of Fig.~\ref{fig3}, we plot $M$ versus 
$r_s$ for $\beta=-5$ with SLy EOS.
We observe that there is a jump from the GR point (a) to 
the VS (b) at $\rho_c=\hat{\rho}_{c1}$. 
This is attributed to the fact that the nonvanishing VS 
suddenly appears as in case (ii) on the left panel 
of Fig.~\ref{fig3}. 

From point (a) to (b), the radius $r_s$ is slightly decreased from 
$11.9$ km to $10.8$ km, but the mass $M$ is increased from 
$0.85M_{\odot}$ to $2.36M_{\odot}$. 
This behavior mostly arises from the nontrivial radial dependence 
of $\rho$. On the VS (b), the coupling $|\beta|$ of order 1 
leads to the increase of $P(r)$ as a function of $r$ around $r=0$, 
which is associated with the decrease of $f(r)$, 
see Eqs.~(\ref{so1}) and (\ref{so3}). 
This is also accompanied by the growth of $\rho(r)$ with $r$ deep inside NS. While $\rho(r)$ starts to decrease around the surface of NSs, the quantity $y=\rho/\rho_0$, 
which appears in Eq.~(\ref{ms}), 
is larger than that in GR in most internal regions of the star. 
This results in the mass $M$ for the VS (b) 
greater than that of the GR point (a). 
For the VS (b), the decrease of $P(r)$ around the surface of star 
occurs more rapidly in comparison to GR, so the radius 
$r_s$ is even smaller than that of point (a).

For $\rho_c$ exceeding $\hat{\rho}_{c1}=3.74 \rho_0$, we observe 
in case (iii) of Fig.~\ref{fig3} that the two intersection points 
with $\bA_{\infty}=0$ start to appear besides the GR branch.
The root with smaller $\bA_c$ (denoted as P$_{s}$) 
has a similar property to the scalarized 
solution in scalar-tensor theories (see Appendix), in that 
the positive shift of $\bA_c$ gives the positive change of $\bA_{\infty}$.
On the other hand, 
the root with larger $\bA_c$ (denoted as P$_{l}$) 
has an opposite characteristic, 
similar to the 0-node solution studied in Sec.~\ref{0node}. 
We also note that, under the shift $\bA_c>0$, the GR point 
in case (iii) moves to the direction $\bA_{\infty}<0$. 
With these properties, there is a possibility 
that the GR branch can be unstable to reach the point P$_s$.
On the other hand, the solution P$_{l}$ should arise from some nonlinear 
effects rather than spontaneous vectorization.

Since the value $\bA_c$ for the root P$_s$ gets smaller 
for increasing $\rho_c$, 
the effect of coupling $\beta$ on modifying the structure of NSs 
tends to be weaker. 
In the right panel of Fig.~\ref{fig3}, we plot the values of 
$M$ and $r_s$ corresponding to the root P$_s$.
As $\rho_c$ increases from point (b), both $M$ and $r_s$ 
approach those in GR. 
Above the critical density $\hat{\rho}_{c2}=7.1 \rho_0$, 
the root P$_{s}$ disappears, see case (iv) in Fig.~\ref{fig3}. 
For $\rho_c>\hat{\rho}_{c2}$, the theoretical curve in the 
$(M,r_s)$ plane is identical to that in GR. 
The difference from the GR solution is present for the density 
in the range $\hat{\rho}_{c1}<\rho_c<\hat{\rho}_{c2}$. 
For $\rho_c$ close to $\hat{\rho}_{c2}$, 
the pressure $P(r)$ decreases as a function of $r$ along 
with the increase of $f(r)$.

In the left panel of Fig.~\ref{fig4}, we plot the mass-radius relation 
for SLy EOS with $\beta=-3, -5, -7$. 
When $\beta=-3$, the root P$_s$ explained above exists for 
the central density $12.32 \rho_0<\rho_c<13.4 \rho_0$. 
In this region, the values of $M$ and $r_s$ are different from those in GR, 
but their modifications are not so significant compared to the 
coupling $\beta=-5$. As $|\beta|$ increases, the 1-node solution  
is present for smaller central densities, e.g., 
$3.74 \rho_0<\rho_c<7.1 \rho_0$ for $\beta=-5$.
When $\beta=-7$, the existence of the 1-node branch is numerically
confirmed even for small $\rho_c$ close to $\rho_0$.
In Fig.~\ref{fig4}, the plotted values of $M$ and $r_s$ 
for $\beta=-7$ correspond to the density region $\rho_c>1.5\rho_0$. 
In this case, the 1-node solution disappears 
above the critical density $\hat{\rho}_{c2}=5.3 \rho_0$. 
Thus, for increasing $|\beta|$, the $M$-$r_s$ relation 
at higher densities is hardly modified in comparison to GR. 
For $|\beta|$ exceeding the order of 10, the 1-node solution 
tends to disappear, but the 2-node solution starts to appear.

The right panel of Fig.~\ref{fig4} shows the theoretical values of 
$M$ and $r_s$ for BSk20 EOS with $\beta=-3.5, -5.0, -6.5$. 
For larger $|\beta|$, the modification from GR occurs at smaller 
central densities in a similar way to  
SLy EOS with $\beta=-{\cal O}(1)$.
Thus, the properties of 1-node solutions as well as 0-node solutions 
are insensitive to the choice of NS EOSs. 
We also compute the gravitational binding energy $\Delta=M_p-M$ for 
1-node solutions and find that the necessary condition $\Delta>0$ 
for gravitational stability is satisfied for the cases plotted in Fig.~\ref{fig4}.

Before closing this section, we would like to discuss whether
the 1-node solutions can be the endpoints of 
tachyonic instability of the GR solutions
with $A_\mu=0$. As we have already seen in Sec.~\ref{consec}, 
the value of $\beta$ for the existence of 1-node solutions
is somewhat consistent with the value for the onset of 
tachyonic instability of the GR solution.
The possibility that the fundamental solution is given 
by 1-node solutions may not be surprising.
For instance, in spherically symmetric Proca stars
the temporal component of  the vector field 
has a single node \cite{Procastar},
while in the scalar boson stars
the scalar field has 0 nodes \cite{Boson}.
However, this would not ensure that 1-node solutions are 
the endpoints of tachyonic instability of the GR star solutions 
in our model.

We have observed several qualitative differences between 
1-node solutions and scalarized solutions in the $M$-$r_s$ relation (see Appendix).
First, in the region of low densities, the 1-node branch in GP theories 
is disconnected to the GR branch in the $M$-$r_s$ diagram,
while the scalarized branch is smoothly connected to the latter.
Spontaneous scalarization occurs via a continuous bifurcation from the GR solution
and may be regarded as a continuous phase transition with the order parameter $\phi$, 
in analogy with spontaneous magnetization in ferromagnetic materials.
On the other hand, it seems more plausible that, even if the 1-node solution is 
realized as the consequence of an instability of the GR solution, 
it may be formed via a mechanism like a first-order phase transition, 
rather than a continuous transition.
Another possibility is that 1-node solutions may be formed from a selected 
choice of initial conditions as in the case of 0-node solutions.
There is also an alternative possibility that vectorized solutions
possess nonzero radial and angular components of the vector field 
satisfying the asymptotic condition $A_\mu (r\to \infty)= 0$. 
In Hellings and Nordtvedt theory, Ref.~\cite{Annulli}
showed that the tachyonic instability of GR solutions arises
for the modes with multipole indices $\ell \geq 1$.
The construction of such solutions is beyond the scope 
of our paper. 

\section{Conclusions}

In this paper, we investigated NS solutions in GP theories 
given by the action (\ref{action}) with the vanishing longitudinal 
vector component ($A_1=0$). 
The deviation from GR arises from the nonvanishing 
temporal vector component $A_0$ in the vicinity of NSs, 
with the asymptotic behavior $A_0 \to 0$ at spatial infinity.
The model (\ref{G4model}) allows for the existence of NS solutions 
with a nontrivial profile of the vector field (\ref{VSso}) 
besides the GR solutions with $A_0=0$ everywhere. 
In addition to relativistic stars with constant density 
$\rho$, we considered SLy and BSk20 EOSs to describe 
the realistic nuclear interaction inside NSs.

In Sec.~\ref{consec}, we first studied the vector-field solution 
for relativistic stars with constant $\rho$ to understand 
its general properties semi-analytically. 
Inside the star, the temporal vector component obeys Eq.~(\ref{A0eq}), 
which possesses the GR branch. The necessary condition for 
the realization of spontaneous vectorization to 
a nonvanishing $A_0$ solution corresponds to $\beta<0$. 
Under the conditions (\ref{apro1}) and (\ref{apro2}), Eq.~(\ref{A0eq}) 
reduces to Eq.~(\ref{A0eq2}), whose solutions inside and outside the 
star are given by Eqs.~(\ref{A0int}) and (\ref{A0inf}) respectively.
However, the approximations (\ref{apro1}) and (\ref{apro2})  
lose their validities for 
compactness of the star ${\cal C}$ of order 0.1. 
In particular, the deviation of metric components $f$ and $h$ from 1 
leads to the decrease of $A_0(r)$ inside the star faster than that 
estimated by Eqs.~(\ref{A0int}). 
This results in the existence of nonvanishing vector-field solutions 
even for the coupling $|\beta|$ smaller than the order $1$.

For $\beta=-{\cal O}(0.1)$, we numerically confirmed the existence of 
0-node solutions where $A_0$ monotonically decreases toward the asymptotic 
value 0 at spatial infinity. In Fig.~\ref{fig1}, 
we observe that the 0-node has a positive scalar charge $Q$.
For $\beta=-{\cal O}(1)$, there exists the 1-node solution where 
$A_0$ crosses 0 once and then approaches 0 as $r \to \infty$.
The 1-node possesses a negative scalar charge. 
The $n$-nodes with $n \geq 2$ only arise for the large 
coupling in the range $\beta < -{\cal O}(10)$.
Although we considered the constant-density star in Sec.~\ref{consec},
these properties are independent of the choice of EOSs.

In Sec.~\ref{0node}, we discussed the property of 0-node solutions 
and the mass-radius relation by considering 
SLy and BSk20 EOSs for $\beta=-{\cal O}(0.1)$.
As we see in the left panel of Fig.~\ref{fig2}, 
below a critical central density $\bar{\rho}_c$, there exists the 
0-node solution with $\bA_c>0$ and $\bA_{\infty}=0$
besides the GR branch.
However, the convex-upward property of theoretical curves in the 
$(\bA_{\infty}, \bA_c)$ plane for the 0-node is different from that of 
scalar-tensor theories in the $(\bar{\phi}_{\infty}, \bar{\phi}_c)$ 
plane (see the left panel of Fig.~\ref{figap}). 
Analogous to the discussion of Ref.~\cite{Annulli}, 
we argue that the GR branch should be stable and hence 
the 0-node solution may arise from some nonlinear effects 
rather than from spontaneous vectorization. 
The coupling $\beta$ works to slow down the decrease of pressure 
inside the star, so the radius $r_s$ and mass $M$ corresponding 
to the 0-node solution are greater than those of the GR branch. 
For increasing $|\beta|$ of order 0.1, the deviation from the GR values of 
$M$ and $r_s$ occurs in the region of lower densities.

In Sec.~\ref{1node}, we showed the existence of 1-node solutions 
for $\beta=-{\cal O}(1)$ with 
SLy and BSk20 EOSs. 
The 1-node suddenly arises above a critical density $\hat{\rho}_{c1}$ 
and disappears above a second critical density $\hat{\rho}_{c2}$. 
For the density in the range $\hat{\rho}_{c1}<\rho_c<\hat{\rho}_{c2}$, there 
are two roots of $\bA_{\infty}=0$, P$_{s}$ and P$_{l}$ 
in the $(\bA_{\infty},\bA_c)$ plane, 
besides the GR branch. The root 
P$_{s}$ has a property 
similar to the scalarized solution in scalar-tensor theories, 
so there is a possibility that the former arises out of 
spontaneous vectorization. 
The mass $M$ corresponding to 
root P$_{s}$ is larger than 
that in GR and, as $\rho_c$ increases toward $\hat{\rho}_{c2}$,
$M$ approaches the GR value. 
For larger $|\beta|$, the deviation of $M$ and $r_s$ from 
those in GR is limited to lower-density regions. 

Here, we would like to emphasize that the existence and qualitative properties of the 0-node 
and 1-node solutions are insensitive to different choices of EOSs. 
As we see in Eq.~(\ref{so4}), the temporal vector component $A_0(r)$ around the center 
of NS is mostly determined by the values of $\rho_c$ and $\beta$.
Since the central pressure $P_c$ does not appear for this solution up to 
the order ${\cal O}(r^4)$, the different choice of EOSs hardly affects the behavior of 
the vector-field profile around the center of NS. On the other hand, 
the iterative solution (\ref{so1}) of $P(r)$ is affected by the value of $P_c$ as well as $\rho_c$ and $\beta$. 
Thus, the radius and mass of NS can be modified by choosing different EOSs, but 
the qualitative behavior of vectorized solutions are insensitive to the change of EOSs.
It is of interest to study observational signatures of these solutions through 
the gravitational wave measurements. 
For vectorized solutions the mass-radius relation differs from 
that in GR, so the tidal Love number of NSs \cite{Flanagan:2007ix,Damour:2009vw,Binnington:2009bb,Hinderer:2009ca} 
is also subject to modifications. This can be potentially tested in the GW 
observations of NS mergers.

It is not clear yet whether 1-node solutions constructed in this paper are 
indeed the endpoints of tachyonic instability of the GR star,
namely, vectorized NS solutions.
Further studies will be needed to clarify this issue.
First, it will be very crucial to investigate whether the 0-node 
and 1-node solutions found in this paper are stable against 
axial and polar perturbations.
Second, it will also be important to construct
NS solutions with the nonvanishing radial and angular components of 
the vector field satisfying $A_\mu (r\to \infty)=0$ and, if they exist, 
check their stability.

On the other hand, the 0-node and 1-node solutions constructed in this paper 
deserve for further studies from various aspects, e.g., the extension to rotating solutions
and the analysis regarding the tidal deformability
and universal relations \cite{Yagi,Yagi2,Yagi3,Stein,ICrel}
(see also \cite{Doneva} and references therein). 
They would be helpful to distinguish the solutions in GP theories from GR and 
other modified theories of gravitation from the theoretical and observational viewpoints.
Another possible issue is that, assuming the vector field triggering 
spontaneous vectorization exists since the beginning of the Universe,
the same coupling can potentially induce tachyonic growth of the vector field 
over the cosmic expansion history, which might result in   
the conflict with Solar System tests of gravity. 
The similar issue has already been pointed out in the case of 
spontaneous scalarization \cite{Cosmology1,Cosmology2,Cosmology3,Cosmology4}.
In vector-tensor theories, the problem may be more serious as the vector field 
has more degrees of freedom and it can also break the isotropy of the Universe.
We hope to come back to these issues in our future work.

\section*{Acknowledgements}

RK is supported by the Grant-in-Aid for Young Scientists B 
of the JSPS No.\,17K14297. 
ST is supported by the Grant-in-Aid for Scientific Research Fund of the 
JSPS No.\,19K03854 and MEXT KAKENHI Grant-in-Aid for Scientific Research 
on Innovative Areas ``Cosmic Acceleration'' (No.\,15H05890).
MM~was supported by the research grant under the Decree-Law 57/2016 of August 29 (Portugal) through the Funda\c{c}\~{a}o para a Ci\^encia e a Tecnologia.
MM~is also grateful for the hospitality at the Tokyo University of Science where this work was initiated.

\appendix

\renewcommand{\theequation}{A.\arabic{equation}}
\setcounter{equation}{0}

\section*{Appendix: Spontaneous scalarization in scalar-tensor theories}

In comparison to the VS in GP theories, we briefly review 
spontaneous scalarization in scalar-tensor theories. 
Let us consider the action in the Jordan frame 
(given by the metric $g_{\mu \nu}$), 
\be
{\cal S}=\int {\rm d}^4 x \sqrt{-g} \left[ \frac{M_{\rm pl}^2}{2} 
F(\phi)R-\frac{1}{2}\omega (\phi) g^{\mu \nu} 
\partial_{\mu} \phi \partial_{\nu} \phi \right]
+{\cal S}_m (g_{\mu \nu}, \Psi_m)\,,
\label{Jaction}
\ee
where $F(\phi)$ and $\omega(\phi)$ are functions of the scalar 
field $\phi$. In this frame, the matter fields are minimally coupled to gravity. 
Under the conformal transformation $(g_{\mu \nu})_E=F(\phi) g_{\mu \nu}$, 
the action (\ref{Jaction}) is transformed to 
\be
{\cal S}
=\int {\rm d}^4 x \sqrt{-g_E} \left[ \frac{M_{\rm pl}^2}{2} 
R_E-\frac{1}{2} (g^{\mu \nu})_E 
\partial_{\mu} \varphi \partial_{\nu} \varphi \right]
+{\cal S}_m \left( F^{-1} (\phi)(g_{\mu \nu})_E, \Psi_m \right)\,,
\label{Eaction}
\ee
where the subscript ``$E$'' represents quantities in the Einstein frame, and 
\be
\frac{\rd \varphi}{\rd \phi}=\sqrt{\frac{3}{2} 
\left( \frac{M_{\rm pl} F_{,\phi}}{F} \right)^2
+\frac{\omega}{F}}\,.
\ee
We choose the canonical scalar field $\varphi$ in the Einstein frame such that 
$\varphi=\phi$. Since $\omega=[1-3M_{\rm pl}^2 F_{,\phi}^2/(2F^2)]F$ 
in this case, the Jordan-frame action (\ref{Jaction}) is expressed as 
\be
{\cal S}=\int {\rm d}^4 x \sqrt{-g} \left[ \frac{M_{\rm pl}^2}{2} 
F(\phi)R-\frac{1}{2} \left( 1-\frac{3M_{\rm pl}^2 F_{,\phi}^2}
{2F^2} \right) F(\phi) g^{\mu \nu} 
\partial_{\mu} \phi \partial_{\nu} \phi \right]
+{\cal S}_m (g_{\mu \nu}, \Psi_m)\,.
\label{Jaction2}
\ee
The nonminimal coupling chosen by Damour and Esposito-Farese \cite{Damour,Damour2}
corresponds to 
\be
F(\phi)=e^{-\beta \phi^2/(2M_{\rm pl}^2)}\,,
\label{Fnon}
\ee
where $\beta$ is a constant.

In full Horndeski theories including the action (\ref{Jaction2}) as a special case, 
the full background equations of motion on the spherically symmetric and 
static background (\ref{metric})  were already derived in the literature, 
see, e.g., Eqs.~(8)-(10) of Ref.~\cite{Kase:2013uja}. 
We do not write them explicitly here.
The pressure $P$ and density $\rho$ obey the same continuity 
equation as (\ref{Peq}). The relation between $P$ and $\rho$ can 
be specified by a given EOS. 
For the nonminimal coupling (\ref{Fnon}), there exists the scalarized 
solution with $\phi(r) \neq 0$ besides the GR branch $\phi(r)=0$. 
On using the expansion similar to Eq.~(\ref{expand}) around $r=0$,
the iterative scalarized solutions to $P$, $h$, $f$, and $\phi$ 
deep inside the star are given by 
\ba
P(r) &=& P_c-\frac{(\rho_c+P_c)[2M_{\rm pl}^2
(\rho_c+3P_c)+\beta^2 \phi_c^2(\rho_c-3P_c)]}
{24M_{\rm pl}^4}e^{\beta \phi_c^2/(2M_{\rm pl}^2)}r^2
+{\cal O} (r^4)\,,\label{Pite}\\
h(r) &=& 1-\frac{2M_{\rm pl}^2 \rho_c-\beta^2 \phi_c^2 
(\rho_c-3P_c)}{6M_{\rm pl}^4}
e^{\beta \phi_c^2/(2M_{\rm pl}^2)}r^2
+{\cal O} (r^4)\,,\\
f(r) &=& f_0+\frac{f_0[2M_{\rm pl}^2
(\rho_c+3P_c)+\beta^2 \phi_c^2(\rho_c-3P_c)]}
{12M_{\rm pl}^4}
e^{\beta \phi_c^2/(2M_{\rm pl}^2)}r^2
+{\cal O} (r^4)\,,\\
\phi(r) &=& \phi_c+\frac{\beta \phi_c (\rho_c-3P_c)}{12M_{\rm pl}^2}
e^{\beta \phi_c^2/(2M_{\rm pl}^2)}r^2
+{\cal O} (r^4)\,,
\label{phiite}
\ea
where $\phi_c$ is the field value at $r=0$. 
At spatial infinity, the scalar field behaves as 
$\phi(r) \simeq \phi_{\infty}+Q/r$, where $\phi_{\infty}$ and $Q$ 
are constants. The internal and external solutions to $\phi(r)$  
are joined each other at the stellar radius $r_s$.
It is known that the 0-node scalarized solution with $\phi_{\infty} \simeq +0$ 
is present for negative $\beta$ in the range 
$\beta<-4.35$ \cite{Harada:1998ge,Novak:1998rk,Silva:2014fca}.

\begin{figure}[h]
\begin{center}
\includegraphics[height=3.2in,width=3.4in]{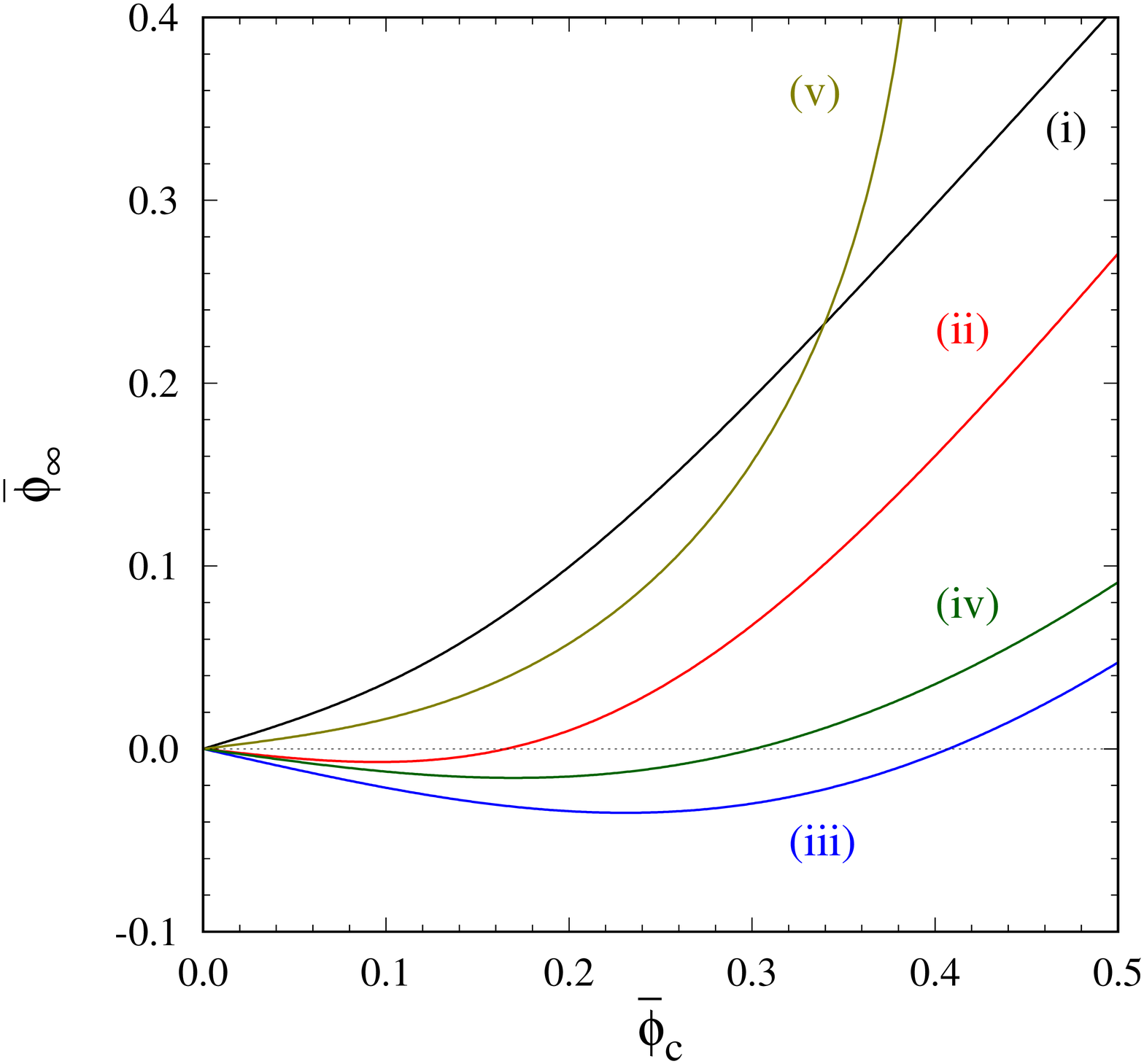}
\includegraphics[height=3.2in,width=3.4in]{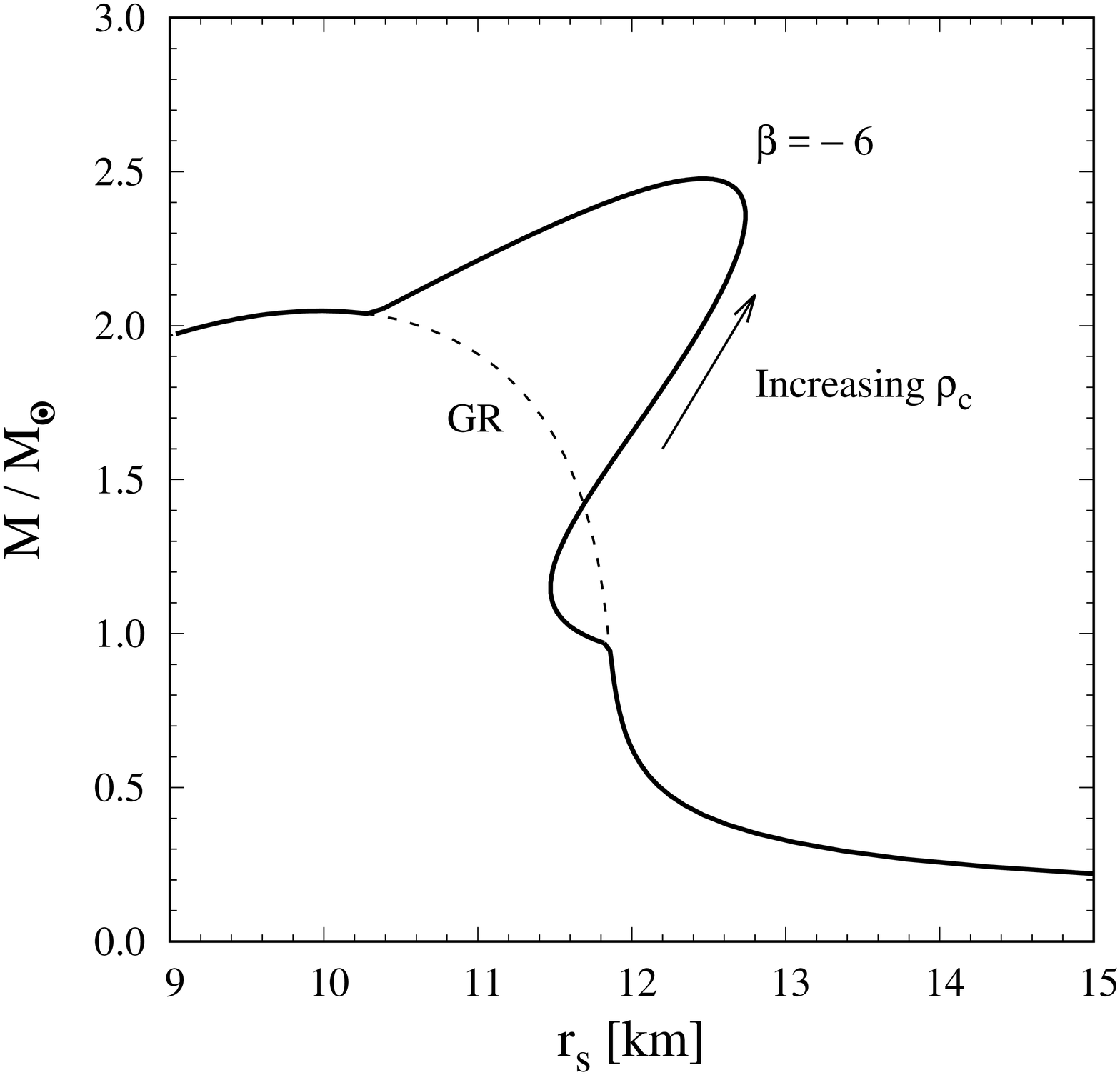}
\end{center}
\caption{\label{figap} 
(Left) 
The scalar field $\bar{\phi}_{\infty}=\phi_{\infty}/M_{\rm pl}$ 
at spatial infinity versus the central value 
$\bar{\phi}_c=\phi_c/M_{\rm pl}$ at $r=0$ 
for SLy EOS with $\beta=-6$. 
Each line corresponds to the central densities 
(i) $\rho_c=3\rho_0$, (ii) $\rho_c=5\rho_0$, 
(iii) $\rho_c=10\rho_0$, (iv) $\rho_c=12\rho_0$, and 
(v) $\rho_c=16\rho_0$, respectively. 
The dashed line represents $\bar{\phi}_{\infty}=0$.
(Right)
$M/M_{\odot}$ versus $r_s$ for SLy EOS with $\beta=-6$. 
We also show how the  values of $M$ and $r_s$  
move with the increase of $\rho_c$.
The mass-radius relation in GR is represented 
as the thin black dashed line.
}
\end{figure}

Let us consider SLy EOS with the coupling $\beta=-6$. 
In the left panel of Fig.~\ref{figap}, we plot
$\bar{\phi}_{\infty}=\phi_{\infty}/M_{\rm pl}$ versus 
$\bar{\phi}_c=\phi_c/M_{\rm pl}$ for five different 
central densities $\rho_c$. 
The line (i), which corresponds to $\rho_c=3\rho_0$, 
has the intersection with $\bar{\phi}_{\infty}=0$ only at the 
GR point $\bar{\phi}_c=0$. 
For $\rho_c \gtrsim 4.3\rho_0$, the scalarized solution 
with $\bar{\phi}_c>0$ and $\bar{\phi}_{\infty}=0$ starts to 
appear besides the GR branch. 
Unlike the 0-node solution in GP theories, 
the line (ii) in Fig.~\ref{figap} is convex downward. 
In this case, the GR solution can be unstable 
to undergo spontaneous scalarization to the other branch 
with $\bar{\phi}_c \neq 0$. 

In the right panel of Fig.~\ref{figap}, we plot 
the mass $M$ and radius $r_s$ of NS for $\beta=-6$, which agrees with 
the result presented in Fig.~1 of Ref.~\cite{Msilva}.
The mass-radius relation is similar to that in GR 
for the central density in the range $\rho_c \lesssim 4.3\rho_0$, but the
difference starts to appear for $\rho_c > 4.3\rho_0$ 
due to the emergence of the scalarized branch.
The radius $r_s$ associated with the scalarized solution 
in the range $4.3\rho_0 < \rho_c  \lesssim 7 \rho_0$ 
is smaller than the corresponding value of the GR branch. 
This is attributed to the fact that $3P_c$ is smaller than $\rho_c$ 
in this regime and hence the term $\beta^2 \phi_c^2(\rho_c-3P_c)$ 
in Eq.~(\ref{Pite}) leads to the decreasing rate of $P(r)$ larger than 
that in GR. As $\rho_c$ increases further, the term $3P_c$
cannot be negligible relative to $\rho_c$.
In particular, for $\rho_c \gtrsim 10\rho_0$, EOS enters 
the fully relativistic region with $\rho_c<3P_c$. 
Then, the term $\beta^2 \phi_c^2(\rho_c-3P_c)$ is negative 
with $e^{\beta \phi_c^2/(2M_{\rm pl}^2)}<1$ for $\phi_c \neq 0$, 
so the decreasing rate of $P(r)$ becomes smaller than that in GR 
deep inside the star. This results in the larger radius $r_s$ 
for the scalarized solution relative to that of the GR branch.
Indeed, the increase of $r_s$ and $M$ seen in Fig.~\ref{figap} 
(in comparison to their GR values)
mostly arises from this slower decrease of $P(r)$.

In the left panel of Fig.~\ref{figap}, we observe that 
the field $\phi_c$ of the scalarized solution reaches the 
maximum value $0.4M_{\rm pl}$ around 
$\rho_c \simeq 10\rho_0$. 
Since $\rho_c-3P_c<0$ for $\rho_c \gtrsim 10\rho_0$, 
the iterative solution (\ref{phiite}) shows that $\phi(r)$ 
increases as a function of $r$ deep inside NSs.
Around the surface of star the term $3P$ becomes smaller 
than $\rho$, so that $\phi(r)$ decreases to 
join the exterior solution at $r=r_s$.
For increasing $\rho_c$, however, this growth of $\phi(r)$ 
tends to occur up to $r=r_s$ and hence it becomes 
more difficult to smoothly connect to the external solution.
Reflecting this point, the field value $\phi_c$ of scalarized 
solutions decreases 
for increasing $\rho_c$ in the range $\rho_c \gtrsim 10\rho_0$, 
see case (iv) of Fig.~\ref{figap}. 
Eventually, the scalarized solution with $\phi_c>0$ disappears 
for $\rho_c > 14.4\rho_0$. The case (v) in Fig.~\ref{figap} 
corresponds to such an example, which possesses only the GR branch. 
In the right panel of Fig.~\ref{figap}, the mass-radius relation
approaches that of GR in this high-density region. 
The above discussion shows that the scalarized solution is present 
in the range $4.3 \rho_0 < \rho_c  < 14.4 \rho_0$ 
for SLy EOS with $\beta=-6$. 
The property of scalarized solutions discussed above also holds for other
EOSs entering the full relativistic regime $\rho_c<3P_c$ 
as $\rho_c$ increases.


\end{document}